\documentclass[journal]{IEEEtran}
\usepackage{graphicx}
\usepackage{subfigure}
\usepackage{multirow}
\usepackage{multicol} 
\usepackage{threeparttable}
\usepackage{arydshln}
\usepackage{booktabs}
\usepackage[backref]{hyperref} 
\usepackage{CJK}
\usepackage{amsmath}
\usepackage{svg}
\usepackage{soul}
\usepackage{changepage}
\ifCLASSINFOpdf
\else
\fi
\begin{document}
%
\title{Sensor Data Augmentation by Resampling for Contrastive Learning for Human Activity Recognition}
%
%
%

\author{Jinqiang Wang,
Tao Zhu,
Jingyuan Gan,
Liming Chen,~\IEEEmembership{Senior Member,~IEEE,}
Huansheng Ning,~\IEEEmembership{Senior Member,~IEEE,}
and Yaping Wan
 
\thanks{Jinqiang Wang, Tao Zhu, Jingyuan Gan and Yaping Wan are with the School of Computer Science, University of South China, 421001 China. e-mail: tzhu@usc.edu.cn.}
\thanks{Liming Chen is with the Ulster University, Northern Ireland, UK. e-mail: l.chen@ulster.ac.uk}
\thanks{Huansheng Ning is with the School of Computer \& Communication Engineering, University of Science and Technology Beijing, 100083 China. e-mail: ninghuansheng@ustb.edu.cn}
}

\maketitle

\begin{abstract}
While deep learning has contributed to the advancement of sensor-based Human Activity Recognition (HAR), it is usually a costly and challenging supervised task with the needs of a large amount of labeled data.  To alleviate this issue, contrastive learning has been applied  for sensor-based HAR. Data augmentation is an essential part of contrastive learning  and has a significant impact on the performance of downstream tasks. However, current popular  augmentation methods do not achieve competitive performance in contrastive learning for sensor-based HAR. 
Motivated by this issue, we propose a new sensor data augmentation method by resampling, which simulates more realistic activity data by varying the sampling frequency to maximize the coverage of the sampling space.  In addition, we extend MoCo, a popular contrastive learning framework, to MoCoHAR for HAR. The resampling augmentation method will be evaluated on two contrastive learning frameworks, SimCLRHAR and MoCoHAR, using UCI-HAR, MotionSensor, and USC-HAD datasets. The experiment results show that the resampling augmentation method outperforms all state-of-the-art methods under a small amount of labeled data, on SimCLRHAR and MoCoHAR, with mean F1-score as the evaluation metric. The results also demonstrate that not all data augmentation methods have positive effects in the contrastive learning framework.
\end{abstract}


%
\IEEEpeerreviewmaketitle

\section{Introduction}
%
%
%
%
\soulregister\cite7
\IEEEPARstart{T}{he} development of technology for sensor-based Human Activity Recognition (HAR)
has brought many intelligent applications into our lives, such as smart homes \cite{rashidi2009keeping},
medical rehabilitation \cite{patel2012review}, \cite{zhou2020deep} and skill assessment \cite{kranz2013mobile}. Due to the popularity of the
Internet of Things, sensors can be better embedded into mobile phones, watches and
other portable devices to obtain a data stream more conveniently. 
Through the analysis and prediction of sensor data by computing systems, computing devices can better understand human behavior, which will play an important role in human health and disease prevention \cite{chen2021deep}.
\par
Currently, there are many methods for processing wearable sensor data (accelerometers, gyroscopes, magnetometers), such as traditional machine learning methods including decision trees, Bayesian networks, and support vector machines \cite{lara2012survey}.
In recent years, deep learning-based methods have been widely used in wearable sensor-based activity recognition tasks. Under supervised learning tasks, models such as LSTM \cite{guan2017ensembles}, CNN \cite{lee2017human}, DeepConvLSTM   \cite{hammerla2016deep},  DeepConvLSTM-Attention \cite{murahari2018attention} and Multi-Head Convolutional Attention \cite{zhang2019novel} have been proposed to significantly improve the accuracy of HAR.
However, this approach 
usually requires a large number of labeled datasets to train a deep learning model, which generally requires manual labeling of sensor data, through a time-consuming and tedious process.
In addition, the labeling are affected by various noise sources, such as sensor noise, segmentation problems, and changes in the activities of different people, which make the annotation process error-prone \cite{chen2021deep}. Therefore, the limitation of sensor data annotation is a major challenge for HAR.
\par
To alleviate the limitations of data annotation, contrastive learning has been proposed as a dominant form of self-supervised learning in computer vision, natural language processing, and other fields \cite{le2020contrastive}.
Contrastive learning generates two different sets of pseudo-labels through data augmentation, enabling the model to distinguish between positive pairs and negative pairs in these two sets. 
By learning  on different augmented versions of the same instance, contrastive learning enables the model to learn the most essential representation of the instance. 
Such a model can perform well in downstream tasks after requiring only a small amount of label data fine-tuning \cite{le2020contrastive}, \cite{jaiswal2021survey}, \cite{falcon2020framework}.
In contrastive learning, data augmentation plays an important role in generating new samples pseudo-labels by simulating different disturbances present in reality \cite{bachman2019learning}.
However, in contrastive learning, current sensor data augmentation methods hardly outperform supervised learning even with a small amount of labels and fail to take full advantage of the features of contrastive learning to improve the activity recognition accuracy \cite{tang2020exploring}.

\begin{figure}

\centering

\subfigure[Raw Sampling Point]{

\label{fig:a} 

\includegraphics[width=8cm]{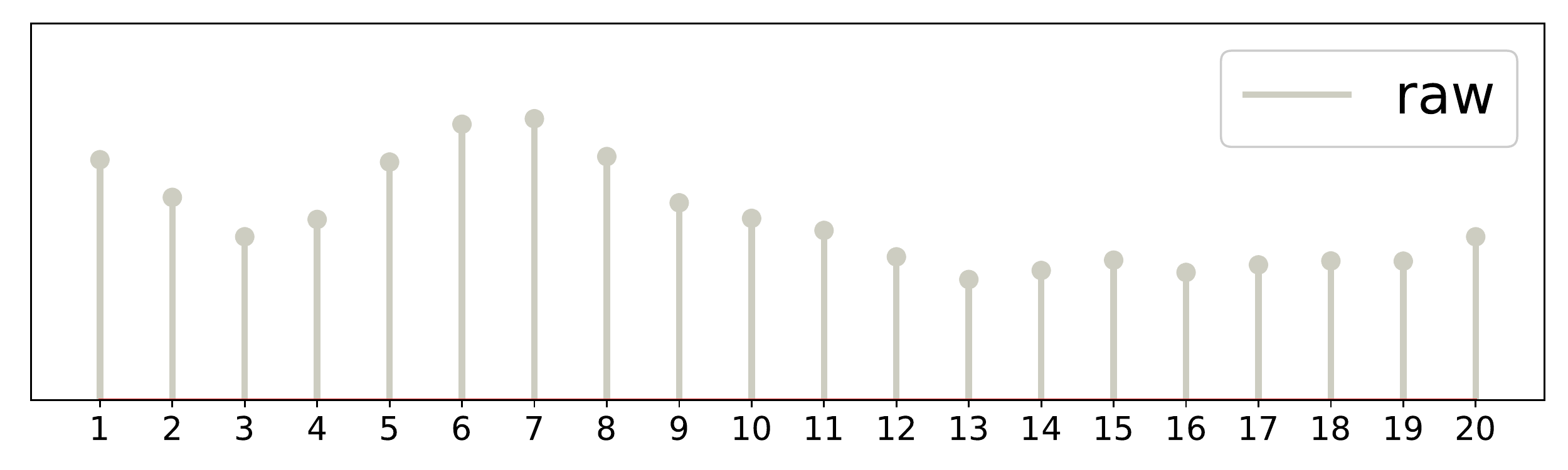}}


\subfigure[Upsampling Point]{

\label{fig:subfig:b} 

\includegraphics[width=8cm]{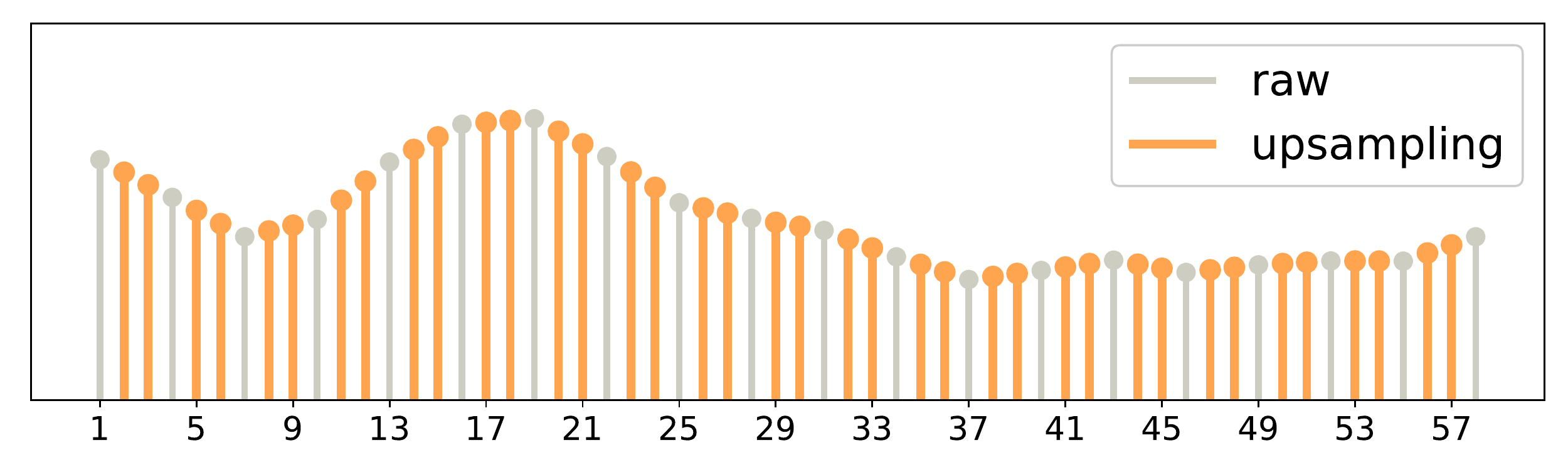}}

\subfigure[Downsampling Point]{

\label{fig:subfig:c} 

\includegraphics[width=8cm]{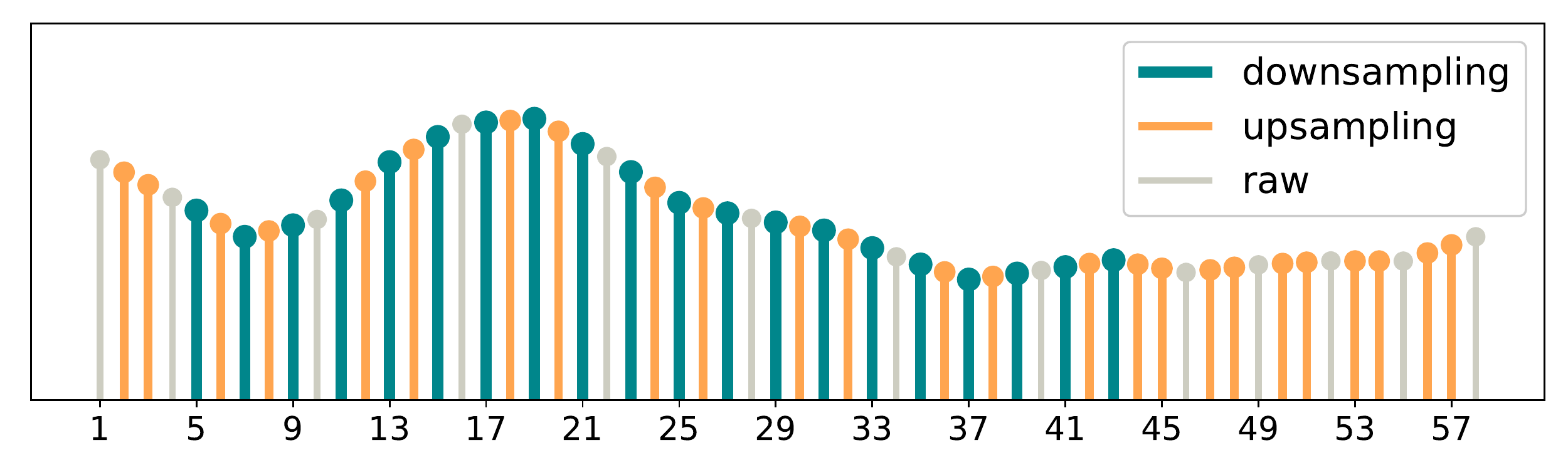}}

\caption{Resampling Diagram}

\label{fig:1} 

\end{figure}
\par
Motivated by the limitations of current augmentation methods, we propose a resampling method for sensor data augmentation. The starting point of this method is to generate more realistic data by varying the sampling frequency to maximize the coverage of the sampling space.  The resampling method is divided into two steps: upsampling and downsampling. Upsampling is the process of fitting new values by using interpolation methods along the sensor data time axis. In this paper, linear interpolation is used unless otherwise specified. Downsampling is the process of filtering values by random or regular sampling to revert to the length of the raw sample. 
A demo will be shown here. Fig. \ref{fig:a} shows the raw sample of  a axis of the acceleration sensor data, which has a total of 20 sampling points.
In the process of upsampling, we used linear interpolation to insert two new sampling points between every two raw sampling points. The orange line in Fig. \ref{fig:subfig:b} is the result after upsampling.
In the downsampling process, we take values at every sample interval until reaching the length of the raw sample. The green line in Fig. \ref{fig:subfig:c} is the result after downsampling.
The whole process of resampling generates green line sampling points from gray line sampling points.
\par
To evaluate the performance of the resampling augmentation method, experiments are first conducted in supervised learning, using DeepConvLSTM as the backbone network, on the UCI-HAR \cite{anguita2013public}, MotionSense \cite{malekzadeh2018protecting}, and USC-HAD \cite{zhang2012usc} datasets, respectively. The experiment results show that resampling outperforms all state-of-the-art augmentation methods at 1\% and 10\% label proportion.
\par
In addition, resampling data augmentation was mainly evaluated on contrastive learning.
We extend SimCLR \cite{chen2020simple} and MoCo \cite{he2020momentum}, \cite{chen2020improved} into HAR-suitable frameworks called SimCLRHAR and MoCoHAR. Note that SimCLRHAR is similar to the work \cite{tang2020exploring}, so it is not included as a contribution point in this paper. The resampling augmentation method will be evaluated in the SimCLRHAR and MoCoHAR contrastive learning frameworks. The experiments were validated on the UCI-HAR, MotionSense, and USC-HAD datasets. The final experiment results show that the resampling augmentation method outperforms state-of-the-art  methods in linear evaluation and fine-tuning under most settings. Finally, we explored the performance of combined augmentation on the final model performance and experiment results found that there are some combined augmentations that outperform the individual augmentations in linear evaluation and fine-tuning.
\par
The contributions of this paper are as follows: 1. A new sensor-based resampling augmentation method is proposed that outperforms other methods in both supervised learning and contrastive learning.
2. We extend MoCo for HAR to include a new resampling data augmentation and DeepConvLSTM encoder, which is called MoCoHAR. This framework is applied to sensor data for the first time and outperforms supervised learning and SimCLRHAR for larger batch sizes.
\par
The remainder of this paper is organized as follows. Section II reviews previous related work and background on sensor data augmentation and self-supervised contrastive learning for HAR. Section III details the resampling data augmentation method and the contrastive learning framework suitable for HAR. In Section IV, to evaluate the performance of the resampling augmentation method, experiment protocols for supervised and contrastive learning were designed. In Section V, experiments with sensor data augmentation methods on supervised learning tasks are presented.
In Section VI, experiments with sensor data augmentation methods on contrastive learning tasks are presented. Section VII summarizes this paper and presents future work based on the identified deficiencies.

\section{background}

\subsection{Contrastive Learning for HAR}
Contrastive learning generates two different sets of pseudo-labels through data augmentation, enabling the model to distinguish between positive pairs and negative pairs in these two sets. Such a model can perform well with a small amount of label fine-tuning in downstream tasks. Thus contrastive learning can alleviate the problem of lack of label data  \cite{jaiswal2021survey}.
The pre-training task of contrastive learning is generally instance discrimination, and its aim is to make different augmented versions of the same instance close to each other, with different instances trying to push apart to obtain the essential features of the raw instance. In computer vision, contrastive learning frameworks represented by SimCLR, MoCo, and BYOL \cite{grill2020bootstrap} have surpassed supervised learning in some datasets for image classification, which shows the great potential of contrastive learning.
\par
At present, a few contrastive learning studies have been applied to HAR.
The study \cite{saeed2020federated} proposes a scalogram contrastive network whose objective at a high level is to contrast raw signals and their corresponding visual representations of the wavelet transform so that a network learns to discriminate between aligned and unaligned scalogram-signal pairs. This approach achieves competitive performance in fully supervised networks and outperforms pre-training using auto encoders in both central and federal contexts. But this approach uses only a single transform methods, which does not capture most of the real interference.
This study \cite{haresamudram2021contrastive} uses Contrastive Predictive Coding \cite{oord2018representation} for the first time in HAR, and it outperforms supervised learning with a small number of labels. However, this method process is more tedious compared to SimCLR and does not outperform supervised learning in linear evaluation.
The study \cite{tang2020exploring} uses a modified SimCLR as a contrastive learning framework, which uses instance discrimination \cite{wu2018unsupervised} as the pre-training task and uses NT-Xent \cite{chen2020simple}, \cite{sohn2016improved} as the loss function.
In the downstream activity classification task, the fine-tuned pretrained model achieved better performance.
This work also analyzed the effects of the combination of different augmentation methods on the activity recognition of contrastive learning, and the best combination of augmentation methods was better than supervised learning.
However,  few combined augmentation methods  outperform supervised learning, and the performance is far less than that of computer vision. It is possible to find an augmentation method that is suitable for sensor data in the framework of contrastive learning to better improve the performance of activity recognition.

\subsection{Sensor Data Augmentation Methods}
For training deep learning models, insufficient labeled data is a major challenge, and data augmentation is a major approach to alleviating this problem. Data augmentation is also critical to the performance of contrastive learning \cite{bachman2019learning}. Data augmentation transforms existing samples into new samples by using a limited amount of data. A key challenge in data augmentation is how to accurately simulate the same class of data under different disturbances. In other words, how to ensure that the augmented samples have the same semantics as the raw samples.
At present, the commonly used sensor data augmentation methods include \cite{um2017data}, \cite{dawar2018data}, \cite{rashid2019times}:
\par
\textbf{Noised}: A method for simulating additional sensor noise by multiplying the raw sample values with values that match a Gaussian or uniform distribution.
\par
\textbf{Rotated}: A method for simulating different sensor positions by plotting a uniformly distributed 3D random axis and a random rotation angle and applying the corresponding rotation to the sample.
\par
\textbf{Scaling}: Multiply by a random scalar to scale the size of the data in the window to simulate the motion of weaker magnitudes.
\par
\textbf{Magnify}: Multiply by a random scalar to magnify the size of the data in the window to simulate stronger amplitude motion.
\par
\textbf{Inverting}: The sample value multiplied by -1 produces a vertical flip or mirror image of the input signal.
\par
\textbf{Reversing}: The entire window of the sample is flipped in the time direction. The second half of the cycle is simulated by the first half of the cycle motion.
\par
\textbf{Permutation}: A simple method to randomly disturb the window temporal positions. The sample is first split into N segments of the same length, and then, the segments are randomly arranged to create a new window.
\par
\textbf{Time warping}: A method for disturbing the temporal position of a sample can use time warping to change the temporal position of the sample by smoothly distorting the time interval between samples.
\par
\textbf{Cropping}: Randomly crop the raw sample according to a certain time window size.
\par
\textbf{Shuffling}: Randomly disrupted channels of sensor data are used to simulate different wearing directions of the sensor.
\par
Most of the above methods are transferred from time series augmentation methods without considering the characteristics of the sensor data,
or the generated new samples cannot well represent the raw label data. Therefore, their performances will fluctuate greatly due to changes in the datasets \cite{um2017data}, \cite{dawar2018data}, \cite{rashid2019times}. The study \cite{tang2020exploring} used the above augmentation methods in contrastive learning, but most of the augmentation methods did not perform better than supervised learning. For this reason, we need to propose an augmentation method that is suitable for contrastive learning.


\section{Methods}
The data augmentation method plays an important role in contrastive learning \cite{bachman2019learning}. To improve the HAR contrastive learning performance, it is necessary to  propose a new augmentation method that is suitable for sensor data.
Data augmentation has the following functions: Increase the amount of training set data to alleviate model overfitting; Simulate natural disturbances to generate more realistic and multi-view data. For example, image augmentation methods can be seen to be simulating different distances, angles, light intensities, tones, positions, and so on. Therefore, the trained model can also be applied to pictures with different light intensities, distances, angles, and positions.
\par
Contrastive learning uses the augmentation method to generate copies of the sample under different disturbances. Then, the model is trained to filter these disturbances by contrastive loss function to make the generated representation better reflect the essential features of the sample.
However, current sensor data augmentation methods do not perform well in contrastive learning \cite{tang2020exploring}, and we need to propose a new augmentation method to address this problem.
\subsection{Resampling Augmentation}
The sensor collects activity data and is sampled at a certain frequency, and thus, a continuous activity signal is usually represented by a discrete sequence of sensor data values.
There can be disturbing factors in the sampling process, such as the sampling frequency, initial time point of sampling (phase angle), noise (such as the movement of the device, interference generated inside the device), and the duration of sampling. 
To avoid the effect of these disturbances on the recognition performance of the model, we need to design augmentation methods to simulate these disturbances. Consequently, the model can learn these disturbances during training to avoid misclassifying sensor data during the inference phase.

To simulate these disturbances and address the limitations of current augmentation methods, we proposed a sensor data augmentation method called resampling, which simulates multiple disturbances by varying the sampling frequency of sensor data and maximizing the coverage of the sampling space.
\par
The resampling method is divided into two stages: upsampling and downsampling. Upsampling generates new sampling points on the time axis by interpolation methods (e.g., linear interpolation, cubic spline interpolation) to simulate the sampling points after increasing the sampling frequency. In this paper, linear interpolation is used for the interpolation methods unless otherwise specified. Down sampling is taken on the time axis according to random or regular sampling to keep the sample length constant. A concrete example is shown in Fig. \ref{fig:1}.
\par
The raw sample data for one axis of the sensor can be expressed as
\begin{equation}
	X[i], i=1,...,I \label{eq1}
\end{equation}

\par
Its values are arranged in chronological order, with a total of I moment data. The upsampling process inserts $M$ equal partition nodes in a linear interpolation approach between two moments, where $M$ is an integer, $M>=1$, and new interpolation nodes can be generated according to Eq. \eqref{eq2}.
The total length of the sequence $X^{'}$ after interpolation is ${L = (M+1) * (I-1)+1}$. At this point, the upsampling process is completed.
\begin{align}
	X^{'}[(M+1)*(i-&1)+k]=X[i]+(X[i+1]-X[i])*\frac{k-1}{M} \notag \\
	&i=1,...,I; k=1,...,M \label{eq2}
\end{align}

\par
After upsampling, the samples must be downsampled to recover the raw time series length. The method takes the value every $N$ time intervals, where $N$ is an integer, $0<=N<=M-1$.
To ensure that the length of the sequence remains unchanged and more approaches to sampling can be taken, the starting point of taking values is selected according to Eq. \eqref{eq3}, where $random(a,b)$ means taking any one value from a to b. After that, the values are taken according to Eq. \eqref{eq4}, and finally, the total length of the sequence $X^{''}$ is I. At this point, the resampling process is completed.
\begin{align}
&s = random(1,L-I*N)\label{eq3}\\
X^{''}[i]=&X^{'}[s+(i-1)*N],{i=1,...,I} \label{eq4}
\end{align}

\subsection{Contrastive Learning for HAR}

Contrastive learning has become a popular unsupervised learning method in the field of computer vision, such as SimCLR and MoCo,  which have a similar structure and both use instance discrimination \cite{wu2018unsupervised} as a pre-training task. The general process of contrastive learning is that the raw samples are first augmented with different methods to obtain two samples under different disturbances, and a dimension-specific representation is obtained after encoding by an encoder. The pre-training task is to distinguish which augmented samples representations are from the same instance and which augmented samples representations are not from the same instance and then to make different augmented representations of the same instance similar and different instances far apart so that to obtain the most essential representation of the raw sample. The model generated by pre-training can better serve the downstream tasks.
\par
To evaluate the performance of resampling augmentation methods in contrastive learning, we follow SimCLR and MoCo, two contrastive learning frameworks applied in computer vision, to construct two contrastive learning frameworks SimCLRHAR and MoCoHAR applied in HAR.

\subsubsection{SimCLRHAR}
\begin{figure*}
  \centering
  \includegraphics[height=1.8in,width=7in]{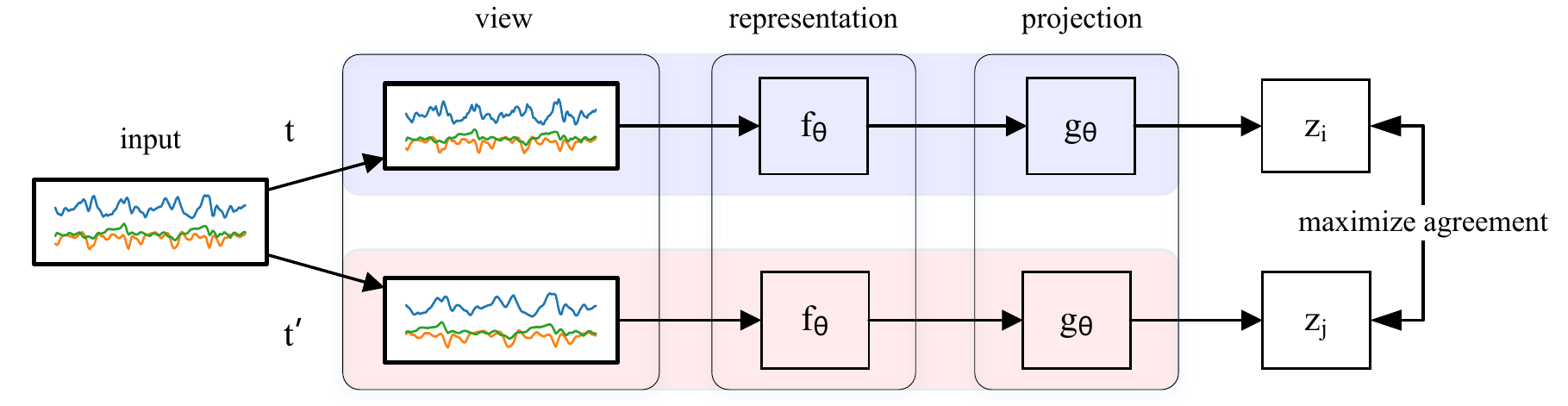}
\caption{SimCLR for Human Activity Recognition (SimCLRHAR)}
  \label{fig:2}
\end{figure*}
SimCLR shows in a simple and intuitive way the general process of contrastive learning in the field of computer vision, where a better result can be obtained by turning up the batch size within a certain range.
In this paper, we extend SimCLR for HAR to include a new resampling data augmentation and DeepConvLSTM encoder, which is called SimCLRHAR. This framework is shown in Fig. \ref{fig:2}.
\par
\textbf{Data Augmentation: }Two different augmentation methods are used for the raw sensor samples to generate two new samples. Alternatively, one branch does not use augmentation and uses the raw sample directly, and the other branch augments.
\par \textbf{Encoder: }Using DeepConvLSTM as the base encoder, which is a classical framework in sensor activity recognition, two newly generated samples are encoded, which in turn generates a representation of certain dimensions. The encoder parameters are shared between the two branches.
\par \textbf{Projection Head: }According to the experience of the work \cite{chen2020simple}, we use a nonlinear projection head to remap the representation generated by the encoder to a new dimension representation. The projection header parameters are shared between the two branches.
\par \textbf{Contrastive Loss Function: } NT-Xent \cite{chen2020simple}, \cite{sohn2016improved} is used as the loss function, and the action objects are $Z_i$ and $Z_j$ generated each time. The loss function formula is as follows:

\begin{equation}
   l(i,j)=-\log {\frac{exp(sim(i,j)/\tau )}{\sum_{k=1}^{2N}I_{[k\neq i]}exp(sim(i,k)/\tau)}} 
   \label{eq5}
\end{equation}
\begin{equation}
   \mathcal{L} = \frac{1}{2N}\sum_{k=1}^{N}[l(2k-1,2k)+l(2k,2k-1)]
   \label{eq6}
\end{equation}
\par where $sim(a,b)$ is the cosine similarity function, $\tau$ denotes a temperature parameter, and N is the batch size. $ I_{[k\neq i]} $ is an indicator function, whose value is 1 when 
k is not equal to i.    
 2k-1 is a positive sample pair with 2k only, and 2k-1 and other values are negative sample pairs.
\par \textbf{Return: }Discard the projection header and return to the encoder.
\par The above is the entire structure of SimCLRHAR, which is similar to the work \cite{tang2020exploring}, and for this reason, it is not included as a contribution point in this paper, but it has new augmentation methods and encoders relative to previous work.
\subsubsection{MoCoHAR}
\begin{figure*}
  \centering
  \includegraphics[height=1.8in,width=7.5in]{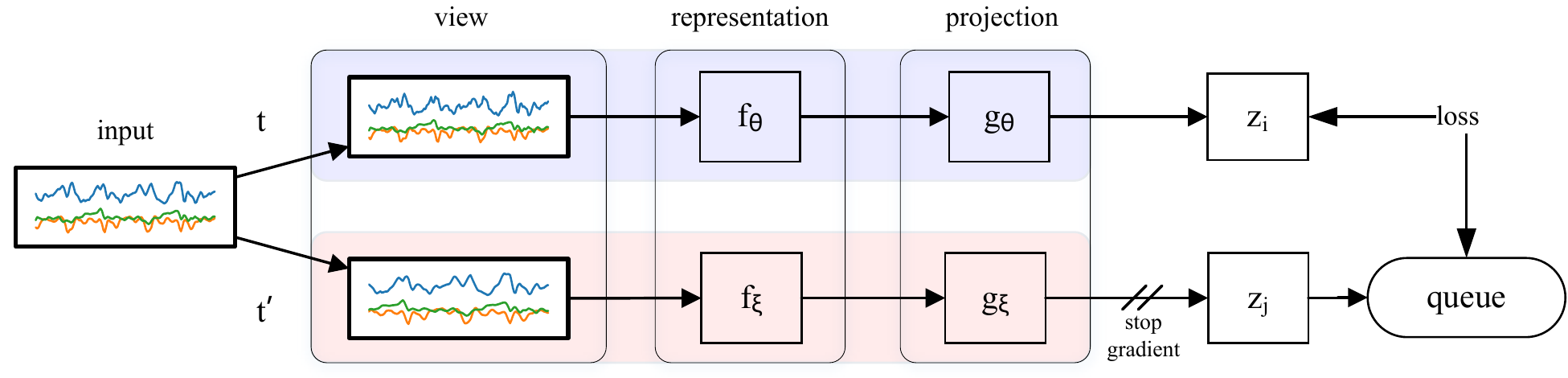}
\caption{MoCo for Human Activity Recognition (MoCoHAR)}
  \label{fig:3}
\end{figure*}

MoCo expands the negative samples for the contrastive approach by using a memory queue, which in turn can obtain better results with a smaller batch size. In this paper, we extend MoCo for HAR to include a new resampling data augmentation and DeepConvLSTM encoder, which is called MoCoHAR. This framework is shown in Fig. \ref{fig:3}.
\par \textbf{Data Augmentation: } Two different augmentation methods are used for the raw sensor samples to generate two new samples. Alternatively, one branch does not use augmentation and uses the raw sample directly, and the other branch augments.
\par \textbf{Encoder and Projection Head: }Two newly generated samples were encoded using DeepConvLSTM as the basic encoder to generate a representation of certain dimensions.
The representation is then remapped to a new dimension using a nonlinear mapping header.
The network parameter $\theta$ is updated by the back propagation of the neural network, and the network parameter $\xi$ is updated in the following manner:
\begin{equation}
    \xi\leftarrow m\xi + (1-m)\theta
    \label{eq7}
\end{equation}
\par Here, $m\in [0,1)$ is the momentum coefficient, the queue shape is $(K, P)$, where $K$ represents the capacity of the queue, and $P$ is the dimension finally generated by the projection head. The $Z_j$ generated each time will be added to the queue. If the queue stack is full, the earliest data added to the queue will be overwritten.
\par \textbf{Contrastive Loss Function: }InfoNCE \cite{oord2018representation} is used as the loss function,
and the action objects are $Z_i$ and the queue generated each time. The loss function formula is as follows:
\begin{equation}
    \mathcal{L}_{q}=-\log{\frac{exp(q\cdot k_{+}/\tau )}{\sum_{i=1}^{K}exp(q\cdot k_{i}/\tau)}}
    \label{eq8}
\end{equation}
\par where $q$ is $Z_i$ generated each time by the network, $ k_+ $ is the positive sample that corresponds to $q$ in the queue, and $\tau$ is the temperature parameter.
\par \textbf{Return: }Return only the encoder for which the network parameter is $\theta$, discarding all other structures.

\section{Experiment}
\subsection{Datasets}
The UCI-HAR \cite{anguita2013public} activity recognition dataset was built from the recordings of 30 subjects performing basic activities and postural transitions while carrying a waist-mounted smartphone with embedded inertial sensors. Six basic activities were included: standing, sitting, lying, walking, upstairs and downstairs. Experiments captured 3-axis linear acceleration and 3-axis angular velocity at a constant 50 Hz rate using the device's built-in accelerometer and gyroscope. In this experiment, 128 readings are sampled as a sliding window, and the sliding window has 50\% overlap.
\par
The MotionSense \cite{malekzadeh2018protecting} dataset consists of time-series data generated by accelerometer and gyroscope sensors. An iPhone 6s was placed in the participant's front pocket, and information was collected from the core motion framework on the IOS device using SensingKit. All of the data was collected at a 50 Hz sampling rate. A total of 24 participants of different genders, ages, weights and heights performed six activities: downstairs, upstairs, walking, jogging, sitting and standing in 15 trials under the same environment and conditions. In this experiment, 200 readings are sampled as a sliding window, and the sliding window has a 12.5\% overlap.
\par
The USC-HAD \cite{zhang2012usc} dataset was collected on the MotionNode sensing platform and contains accelerometer and gyroscope data. The dataset consists of data from 14 subjects, which recorded 12 activities, including walking forward, walking left, walking right, going upstairs, going downstairs, running forward, jumping, sitting, standing, sleeping, going up and going down the elevator. All data were collected at a 100 Hz sampling rate. In this experiment, 200 readings are sampled as a sliding window, and the sliding window has a 25\% overlap.
\par
The experiments in this paper will use the accelerometer and gyroscope data from the above datasets.
\subsection{Data Augmentation}
Based on the research \cite{tang2020exploring}, \cite{saeed2019multi}, this paper adopts the following parameter setting augmentation method:
\par \textbf{Noised: }Add random noise signals with a maximum value of 0.1 and a minimum value of -0.1 to the data sample, which are subject to a uniform distribution.
\par \textbf{Rotated: }Draw a uniformly distributed three-dimensional random axis and a random rotation angle, and apply the corresponding rotation to the sample, rotating according to the random angle.
\par \textbf{Scaling: }Each channel of the signal is scaled by a random value between 0.7 and 0.9.
\par \textbf{Magnify: }Each channel of the signal is amplified by a random value between 1.1 and 1.3.
\par \textbf{Inverting: }The sample value times negative 1.
\par \textbf{Reversing: }The entire window of the sample is flipped in the time direction.
\par \textbf{Resampling: }Upsampling involves inserting a specified or random number of new sampling points between every two sampling points, and the interpolation method is linear. Downsampling takes the value with every specified or random time interval.
\par Due to the poor performance of the other augmentation methods in the work \cite{tang2020exploring}, the experiment in this paper no longer compares them.

\subsection{Supervised Learning Experiment Setup}

\par
The performance of the resampling augmentation method is first validated on supervised learning, and the experiment backbone network is DeepConvLSTM, which is tested on the UCI-HAR, MotionSense, and USC-HAD datasets. To simulate different degrees of labeled data shortages, the training set proportions were set to 1\%, 10\%, and 60\%. The code is built on the TensorFlow platform with an optimizer using Adam \cite{kingma2014adam} and an initial learning rate of 5e-4. The batch size is set to 50, 500, and 1000 according to the different training set proportions of 1\%, 10\%, and 60\%, respectively. The model was trained for 200 epochs, using mean F1-score as the evaluation metric. An NVIDIA Tesla V100 GPU was used to accelerate the training process. All of the experiments following this part of the setup were trained 10 times in different training and test sets divided, and the test results were averaged.
\subsection{Contrastive Learning Experiment Setup}
In this paper, two contrastive learning frameworks, SimCLRHAR and MoCoHAR, are used to evaluate the performance of the resampling augmentation method on HAR.
\par \textbf{SimCLRHAR: }The pre-training task uses DeepConvLSTM as the backbone network and NT-Xent as the contrastive loss function. Three layers of the MLP projection head are added with dimensions 256, 128, and 50. The optimizer uses Adam with an initial learning rate of 1e-3. The temperature is 0.1, the batch size is 2048, and 200 epochs are trained. This pre-training uses all of the data from a single dataset.
\par \textbf{MoCoHAR: }The pre-training task uses DeepConvLSTM as the backbone network and InfoNCE as the contrastive loss function. Two layers of MLP projection heads with dimensions of 256,128 are added. The optimizer is Adam with an initial learning rate of 1e-3. The temperature is 0.07, K is 8192, m is 0.999, the batch size is 1024, and 200 epochs are trained. This pre-training uses all of the data from a single dataset.
\par After pre-training, the encoder is obtained. To verify the performance of the pre-training model, this paper adopts two evaluation protocols to evaluate the downstream classification tasks:
\par Linear evaluation:
Freeze the encoder and add a linear classification layer at the end of the model.
The optimizer uses Adam, and the initial learning rate is 1e-2.
\par Fine tuning: Unfreeze the encoder and add a linear classification layer at the end of the model.
The optimizer uses Adam, and the initial learning rate is 5e-4.
\par In both evaluation experiments, the loss function is cross-entropy loss. The batch size is set to 50, 500, and 1000 according to different training set proportions of 1\%, 10\%, and 60\%, respectively. The model was trained for 200 epochs, using mean F1-score as the evaluation metric. All of the experiments were trained 10 times in different training and test sets divided, and the test results were averaged.

\section{Sensor Data Augmentation Methods for Supervised Learning}
\subsection{Resampling Hyperparameter Study}
To explore the best performance and parameter sensitivity of the resampling augmentation method, two different values of the hyperparameters are listed: 1. $M$ new nodes are inserted between every two moments; 2. A value is taken every $N$ time intervals. Before training, the training set samples are augmented four times.
The specific experiment results are shown in Table \ref{Tab01}.

\begin{table}[]

\scriptsize

\centering

\caption{Hyperparametric sensitivity analysis}

\label{Tab01}

\begin{tabular}{ccccccc}

\toprule

\multirow{2}{*}{} & \multicolumn{1}{c}{$M=1$} & \multicolumn{2}{c}{$M=2$} & \multicolumn{3}{c}{$M=3$} \\

\cmidrule(r){2-2} \cmidrule(r){3-4} \cmidrule(r){5-7}

&  $N=0$  

&  $N=0$      &  $N=1$   

&  $N=0$      &  $N=1$  &   $N=2$\\

\midrule

UCI-HAR             & \textbf{88.20}                   & 86.63           & 87.76           & 87.14          & 88.01           & 86.89      \\

MotionSense              & 87.05                  & 86.56           & \textbf{87.20}           & 87.02          & 86.95           & 86.52             \\
USC-HAD &71.09   &69.37  &70.85  &68.51  &\textbf{71.63}  &70.89\\
\bottomrule

\end{tabular}

\end{table}

\par
The experiment results show that the values of different parameters of the resampling augmentation method can affect the classification performance of the supervised task to varying degrees. In other words, the results of this experiment demonstrate that different sampling frequencies can impact activity classification performance. This finding is caused by the fact that when humans perform the same movement, the amplitude of the movement we use each time is not exactly the same. For example, when walking, the amplitude of our swinging arms and the span of our legs will be different, and thus, the data collected by the sensors will also be different, and we simulate changing the sampling frequency of the sensors, which is equivalent to generating one more set of the subjects' motion data that is closer to the sample of that subject in the test set.

\subsection{Nonlinear Interpolation Study}
There are many nonlinear interpolation methods, such as Lagrange interpolation and cubic spline interpolation.
In this subsection, we will explore the performance of replacing linear interpolation with Lagrange interpolation \cite{waring1779vii} and cubic spline interpolation \cite{mckinley1998cubic} in the upsampling phase. We use segmented nonlinear interpolation in the upsampling phase and a randomly clipped segment of continuous samples in the downsampling phase. The experiment results are shown in Table \ref{Tab02}.
\begin{table*}[]
  \centering
  \caption{nonlinear interpolation}
  \label{Tab02}
  \begin{tabular}[]{cccccccccc}
    \toprule
    \multirow{2}{*}{Mode} & \multicolumn{3}{c}{UCI-HAR} & \multicolumn{3}{c}{MotionSense}& \multicolumn{3}{c}{USC-HAD}\\
    \cmidrule(r){2-4} \cmidrule(r){5-7} \cmidrule(r){8-10}
     &Linear  &Lagrange &Cubic Spline & Linear  &Lagrange &Cubic Spline& Linear  &Lagrange &Cubic Spline\\
    \midrule
    A  &-  &86.16  &85.42	&-	&86.35	&86.32      &- &70.83 &71.03\\
    B  &-  &85.45  &87.25	&-	&87.35	&87.24      &- &71.16 &71.02\\
    Best  &88.20  &86.16  &87.25	&87.20	&87.35	&87.24 &71.63 &71.16 &71.03\\
    \bottomrule
  \end{tabular}
  \begin{tablenotes}
  	\centering
    \footnotesize
    \item[1] Mode A is to fit a curve for every four consecutive sample points in each channel of sensor data, inserting a sample point between the 2nd and 3rd sample points.  
    Mode B is to fit a curve for every eight consecutive sampling points, inserting one sample point between the 2nd and 3rd sampling points, the 4th and 5th sampling points, and the 6th and 7th sampling points, respectively.
  \end{tablenotes}
\end{table*}
\par The experiment results show that linear interpolation performs better than nonlinear interpolation in the UCI-HAR and USC-HAD datasets. In the MotionSense dataset, linear interpolation and nonlinear interpolation perform very similarly. In general, multiple interpolation methods perform similarly. We analyze that no matter which interpolation method is used, they aim to change the sampling frequency. Any interpolation method that conforms to 
the resampling will likely produce a similar performance. This finding demonstrates also that the upsampling phase of the resampling method is suitable for multiple interpolation methods.
\subsection{Comparison with State-of-the-art augmentation methods under supervised learning}
To evaluate the performance of resampling augmentation methods on supervised tasks, we compared other existing sensor data augmentation methods. The final experiment results and 95\% confidence limits are shown in Table \ref{Tab03}, while we performed Wilcoxon signed-rank test \cite{woolson2007wilcoxon} by resampling method with other methods, and the non-parametric statistical hypothesis test result is shown in Table \ref{Tab04}.
\begin{table*}
  \centering
  \caption{Comparison with State-of-the-art augmentation methods under supervised learning}
  \label{Tab03}
  \setlength{\tabcolsep}{1mm}{
  \begin{tabular}[]{cccccccccc}
    \toprule
    \multirow{2}{*}{} &\multicolumn{3}{c}{UCI-HAR} & \multicolumn{3}{c}{MotionSense} & \multicolumn{3}{c}{USC-HAD}\\
    \cmidrule(r){2-4} \cmidrule(r){5-7} \cmidrule(r){8-10}
    &1\%	&10\%	&60\%	&1\%	&10\%	&60\% &1\%	&10\%	&60\%\\
      
    \midrule
    \multirow{2}{*}{Supervised}  &77.94&94.46&96.13     &81.13&96.23&98.62  &63.93&85.58&90.94\\
    &[75.38,80.50]&[94.30,94.62]&[95.78,96.48]  &[79.02,83.24]&[96.01,96.46]&[98.49,98.74]  &[62.28,65.58]&[85.22,85.93]&[90.78,91.10]\\
    \multirow{2}{*}{Noise} &80.14&95.12&97.95   &80.14&96.55&99.08      &65.31&85.77&90.52\\
    &[77.93,82.35]&[94.87,95.36]&[97.55,98.35]  &[78.44,81.83]&[96.43,97.04]&[98.97,99.19]  &[63.67,66.94]&[85.58,85.97]&[90.34,90.70]\\
    \multirow{2}{*}{Rotated} &76.52&93.22&95.48     &78.29&94.04&98.13      &53.78&81.22&87.66\\
    &[75.37,77.67]&[92.94,93.49]&[95.08,95.87]  &[76.87,79.72]&[94.66,95.75]&[98.02,98.25]  &[52.36,55.19]&[80.77,81.68]&[87.40,87.92]\\
    \multirow{2}{*}{Scaling} &81.55&94.91&98.23     &81.97&95.73&98.86      &64.94&85.36&91.00\\
    &[79.33,83.78]&[94.68,95.14]&[97.65,98.82]  &[80.68,83.25]&[95.68,96.31]&[98.74,98.97]  &[63.37,66.51]&[85.09,85.63]&[90.82,91.18]\\
    \multirow{2}{*}{Magnify} &83.61&95.11&\textbf{98.39}   &81.54&95.86&98.88      &63.78&85.78&90.99\\
    &[82.36,84.87]&[94.84,95.39]&[98.17,98.60]  &[80.57,82.51]&[95.69,96.49]&[98.64,99.12]  &[62.55,65.02]&[85.41,86.14]&[90.74,91.25]\\
    \multirow{2}{*}{Inverting} &81.60 &94.69&97.66      &77.66&95.82&98.64      &56.91&83.54&90.68\\
    &[79.29,83.90]&[94.33,95.06]&[97.42,97.91]  &[76.23,77.59]&[96.30,96.83]&[98.53,98.74]  &[55.17,58.64]&[82.93,84.15]&[90.41,90.95]\\
    \multirow{2}{*}{Reversing} &84.02&94.81&98.3    &86.28&96.67&98.92      &67.80&85.72&91.49\\
    &[82.38,85.65]&[94.43,95.19]&[97.87,98.73]  &[85.18,87.60]&[96.77,97.03]&[98.78,99.05]  &[66.70,68.89]&[85.20,85.94]&[91.29,91.68]\\
    \multirow{2}{*}{Resampling(ours)}  &\textbf{88.20} &\textbf{95.27}&98.28&       \textbf{87.20}&\textbf{97.00}	&\textbf{99.35}         &\textbf{71.63}&\textbf{86.97}&\textbf{92.28}\\
    &[87.59,88.82]&[95.02,95.51]&[97.79,98.76]  &[86.75,87.65]&[97.21,97.58]&[99.28,99.42]  &[70.54,72.71]&[86.77,87.17]&[92.07,92.48]\\
    \bottomrule
  \end{tabular}
  }
\end{table*}
\begin{table*}
  \centering
  \caption{The statistical analysis results of the resampling augmentation method with other methods about the Wilcoxon signed-rank test on supervised tasks}
  \label{Tab04}
  \setlength{\tabcolsep}{5mm}{
  \begin{tabular}[]{cccccccccc}
    \toprule
    \multirow{2}{*}{} &\multicolumn{3}{c}{UCI-HAR} & \multicolumn{3}{c}{MotionSense} & \multicolumn{3}{c}{USC-HAD}\\
    \cmidrule(r){2-4} \cmidrule(r){5-7} \cmidrule(r){8-10}
    &1\%	&10\%	&60\%	&1\%	&10\%	&60\% &1\%	&10\%	&60\%\\
      
    \midrule
    Supervised  &s+&s+&s+   &s+&s+&s+   &s+&s+&s+\\
  
    Noise       &s+&+&+     &s+&+&s+   &s+&s+&s+\\
   
    Rotated     &s+&s+&s+   &s+&s+&s+   &s+&s+&s+\\
   
    Scaling     &s+&+&+     &s+&+&s+   &s+&s+&s+\\
   
    Magnify     &s+&+&-     &s+&s+&s+   &s+&s+&s+\\
   
    Inverting   &s+&+&s+    &s+&s+&s+   &s+&s+&s+\\
    
    Reversing   &s+&s+&-    &+&+&s+   &s+&s+&s+\\

    \bottomrule
  \end{tabular}
  }
  \begin{tablenotes}
  	\centering
    \footnotesize
    \item All non-parametric statistical hypothesis test results have been shown here at the 0.05 level of significance. These signs "+", "-", "s+", "s-" indicate that the resampling method is insignificantly better than, insignificantly worse than, significantly better than, and significantly worse than other augmentation methods, respectively.
  \end{tablenotes}
\end{table*}
\par The experiment results show that the resampling method outperforms the supervised learning and all state-of-the-art data augmentation methods for classification performances with 1\% and 10\% labeled data. In particular, the resampling augmentation method improves significantly under 1\% labeled data, outperforming the best method by 4.18\%, 0.92\%, and 3.83\% in the UCI-HAR, MotionSensor, and USC-HAD datasets, respectively. At 60\% labeled data, the resampling augmentation method outperforms other augmentation methods at the 95\% confidence limit on the MotionSensor and USC-HAD datasets. 
The non-parametric statistical hypothesis test found that the resampling method was significantly better than state-of-the-art augmentation methods in most settings.  Even when the resampling method is worse than the other methods, it is insignificantly worse. This finding demonstrates that the resampling augmentation method can alleviate the lack of labeled data and significantly improve activity classification performance relative to previous methods.

\section{Sensor data augmentation for contrastive learning}
\subsection{Comparison with State-of-the-art augmentation methods under contrastive learning}
To evaluate the performance of the resampling augmentation method, it will be used with two contrastive learning frameworks, SimCLRHAR and MoCoHAR, and it will be compared with other augmentation methods at different labeled data proportions.
To compare more clearly with other augmentation methods, we do not take augmentation in the augmentation phase for the first branch but use the raw samples, and we use specific augmentation methods for the second branch.
The experiment results are shown in Table \ref{Tab05}, while we performed Wilcoxon signed-rank test by resampling method with other methods, and the non-parametric statistical hypothesis test result is shown in Table \ref{Tab06}.

\begin{table*}[]

\scriptsize

\centering

\caption{Comparison with State-of-the-art augmentation methods under contrastive learning}

\label{Tab05}

 \setlength{\tabcolsep}{0.3mm}{
	\begin{tabular}{ccccccccccccc}
		
		\toprule
		
		\multirow{3}{*}{} 
		&\multicolumn{4}{c}{1\%} &\multicolumn{4}{c}{10\%} &\multicolumn{4}{c}{60\%}\\
		\cmidrule(r){2-5}\cmidrule(r){6-9}\cmidrule(r){10-13}
		& \multicolumn{2}{c}{Linear evaluation} & \multicolumn{2}{c}{Fine-tuned} 
		& \multicolumn{2}{c}{Linear evaluation} & \multicolumn{2}{c}{Fine-tuned}
		& \multicolumn{2}{c}{Linear evaluation} & \multicolumn{2}{c}{Fine-tuned}\\
		
		\cmidrule(r){2-3} \cmidrule(r){4-5}\cmidrule(r){6-7} \cmidrule(r){8-9}
		\cmidrule(r){10-11} \cmidrule(r){12-13}
		
		&SimCLR. & MoCo.  &SimCLR. &MoCo.&SimCLR. & MoCo.  &SimCLR. &MoCo.
		&SimCLR. & MoCo.  &SimCLR. &MoCo.\\
		\midrule
		\multirow{2}{*}{UCI-HAR(Sup.)}
		 & \multicolumn{4}{c}{77.94}& \multicolumn{4}{c}{94.46}& \multicolumn{4}{c}{96.13}\\
		 & \multicolumn{4}{c}{[75.38,80.50]}& \multicolumn{4}{c}{[94.30,94.62]}& \multicolumn{4}{c}{[95.78,96.48]}\\
		\multirow{2}{*}{Noise} &54.60&61.63&81.90&82.02     &77.19&76.30&94.63&95.02	&81.82&82.63&97.30&97.25\\
		&[54.30,54.90]&[61.53,61.72]&[80.49,83.32]&[81.03,83.01]    &[76.77,77.60]&[75.85,76.75]&[94.45,94.81]&[94.86,95.18]    &[81.58,82.05]&[82.19,83.06]&[96.90,97.69]&[96.73,97.77]\\
		\multirow{2}{*}{Rotated} &53.72&62.39&79.93&80.89   &72.14&78.81&94.69&94.68	&77.50&84.23&\textbf{97.35}&96.71\\
		&[53.26,54.18]&[62.18,62.59]&[78.85,81.00]&[80.10,81.68]    &[71.76,72.52]&[78.45,79.18]&[94.60,94.78]&[94.56,94.81]    &[77.15,77.84]&[83.95,84.50]&[96.88,97.82]&[96.21,97.22]\\
		\multirow{2}{*}{Scaling} &53.80&62.70&77.98&81.72	&75.67&77.73&94.61&95.08	&79.65&83.31&97.21&\textbf{97.49}\\
		&[53.60,53.99]&[62.47,62.93]&[76.51,79.46]&[80.48,82.96]    &[75.35,75.99]&[77.49,77.99]&[94.41,94.81]&[94.79,95.36]   &[79.40,79.90]&[82.91,83.71]&[96.58,97.84]&[97.13,97.84]\\
		\multirow{2}{*}{Magnify} &57.89&61.81&79.24&80.51	&78.68&75.83&94.49&95.04	&81.76&81.61&96.57&96.79\\
		&[57.68,58.09]&[61.64,61.99]&[77.55,80.93]&[79.58,81.44]    &[78.38,78.97]&[75.59,76.07]&[94.33,94.66]&[94.88,95.20]    &[81.45,82.07]&[81.22,82.01]&[96.14,96.99]&[96.32,97.24]\\
		\multirow{2}{*}{Inverting} &57.70&63.81&73.97&81.11	    &67.04&76.23&94.37&94.77	&70.34&81.75&96.40&96.24\\
		&[57.39,58.02]&[63.58,64.04]&[73.07,74.88]&[80.20,82.02]    &[66.76,67.31]&[75.98,76.49]&[94.07,94.67]&[94.61,94.93]    &[70.18,70.50]&[81.43,82.06]&[95.88,96.92]&[95.96,96.52]\\
		\multirow{2}{*}{Reversing} &63.60&75.30&84.37&83.54	    &75.49&88.31&95.13&95.29	&78.60&90.95&96.85&97.28\\
		&[63.26,63.93]&[74.94,75.65]&[82.87,84.91]&[82.06,85.02]    &[75.22,75.77]&[88.01,88.60]&[94.96,95.30]&[95.09,95.50]    &[78.36,78.85]&[90.77,91.14]&[96.53,97.16]&[96.77,97.78]\\
		\multirow{2}{*}{Resampling(ours)}  &\textbf{67.87}&\textbf{83.56}&\textbf{87.82}&\textbf{87.41}	    &\textbf{80.21}&\textbf{91.89}&\textbf{95.26}&\textbf{95.49}	&\textbf{85.24}&\textbf{93.95}&96.78&96.86\\
		&[67.47,68.26]&[83.36,83.76]&[87.36,88.28]&[86.65,88.17]    &[79.93,80.50]&[91.73,92.05]&[95.07,95.45]&[95.33,95.65]    &[84.87,85.62]&[93.79,94.11]&[96.43,97.12]&[96.35,97.36]\\
		\midrule
		\multirow{2}{*}{MotionSense(Sup.)}& \multicolumn{4}{c}{81.13}& \multicolumn{4}{c}{96.23}& \multicolumn{4}{c}{98.62}\\
		& \multicolumn{4}{c}{[79.02,83.24]}& \multicolumn{4}{c}{[96.01,96.46]}& \multicolumn{4}{c}{[98.49,98.74]}\\
		\multirow{2}{*}{Noise} &51.70&61.86&78.09&80.92	    &73.80&73.36&95.54&96.24	&78.66&78.60&98.62&98.98\\
		&[51.59,51.80]&[61.78,61.95]&[77.08,79.10]&[80.17,81.67]    &[73.55,74.06]&[72.99,73.72]&[94.71,96.37]&[95.97,96.51]    &[78.35,78.96]&[78.42,78.78]&[98.31,98.93]&[98.89,99.08]\\
		\multirow{2}{*}{Rotated} &48.67&63.27&78.67&81.77	&72.53&76.13&95.76&96.16	&76.55&81.66&98.59&98.90\\
		&[48.47,48.88]&[63.21,63.34]&[76.76,80.59]&[80.98,82.55]    &[72.30,72.77]&[75.67,76.58]&[95.54,95.98]&[95.85,96.47]    &[76.29,76.80]&[81.51,81.81]&[98.49,98.68]&[98.78,99.03]\\
		\multirow{2}{*}{Scaling} &48.51&60.23&75.91&81.10	&70.16&73.36&95.37&96.37	&75.35&78.22&98.47&98.87\\
		&[48.41,48.60]&[60.15,60.32]&[74.27,77.56]&[79.82,82.37]    &[69.94,70.38]&[73.05,73.67]&[94.74,96.01]&[96.10,96.65]    &[75.02,75.68]&[77.98,78.46]&[98.32,98.62]&[98.64,99.09]\\
		\multirow{2}{*}{Magnify} &56.25&61.76&77.66&79.73	&69.13&71.61&95.78&\textbf{96.48}	&75.43&78.91&98.64&98.95\\
		&[56.00,56.50]&[61.70,61.81]&[75.99,79.32]&[79.17,80.29]    &[68.90,69.36]&[71.29,71.92]&[95.50,96.06]&[96.25,96.71]    &[75.06,75.79]&[78.76,79.07]&[98.53,98.74]&[98.73,99.17]\\
		\multirow{2}{*}{Inverting} &54.09&62.23&74.79&81.74	    &67.73&72.68&95.27&95.80	&70.67&77.26&98.52&98.80\\
		&[54.04,54.15]&[62.16,62.31]&[72.86,76.71]&[80.84,82.63]    &[67.54,67.92]&[72.14,73.22]&[94.91,95.63]&[95.49,96.11]    &[70.44,70.90]&[76.97,77.54]&[98.19,98.85]&[98.68,98.91]\\
		\multirow{2}{*}{Reversing} &66.99&77.52&83.17&84.89	    &77.58&85.24&96.77&95.39	&81.20&88.02&98.94&98.77\\
		&[66.88,67.11]&[77.42,77.62]&[81.46,84.87]&[83.83,85.94]    &[77.42,77.74]&[85.06,85.43]&[96.58,96.96]&[95.03,95.75]    &[80.86,81.53]&[87.82,88.22]&[98.84,99.04]&[98.67,98.86]\\
		\multirow{2}{*}{Resampling(ours)}  &\textbf{76.45}&\textbf{77.66}&\textbf{86.04}&\textbf{85.48} 	&\textbf{85.84}&\textbf{91.55}&\textbf{97.05}&96.32	&\textbf{89.43}&\textbf{93.64}&\textbf{99.11}&\textbf{99.09}\\
		&[76.32,76.57]&[77.56,77.74]&[84.41,87.66]&[84.60,86.36]    &[85.64,86.06]&[91.40,91.69]&[96.95,97.15]&[96.05,96.59]    &[89.19,89.66]&[93.46,93.81]&[99.04,99.18]&[99.00,99.18]\\
		\midrule
		\multirow{2}{*}{USC-HAD(Sup.)}& \multicolumn{4}{c}{63.93}& \multicolumn{4}{c}{85.58}& \multicolumn{4}{c}{90.94}\\
		& \multicolumn{4}{c}{[62.28,65.58]}& \multicolumn{4}{c}{[85.22,85.93]}& \multicolumn{4}{c}{[90.78,91.10]}\\
		\multirow{2}{*}{Noise}     &36.60&39.33&46.53&61.17     &51.39&50.47&77.87&85.66	&56.54&56.38&88.83&91.20\\
		&[35.93,37.27]&[38.54,40.13]&[44.94,48.12]&[59.47,62.87]    &[50.84,51.39]&[49.84,51.10]&[77.04,78.70]&[85.36,85.97]    &[56.36,56.71]&[56.11,56.65]&[88.67,88.99]&[90.83,91.20]\\
		\multirow{2}{*}{Rotated}   &28.50&42.16&44.58&56.83	    &38.90&55.02&83.58&84.43	&43.26&60.65&89.85&90.52\\
		&[27.77,29.22]&[41.04,43.27]&[42.23,46.93]&[54.73,58.94]    &[38.60,39.21]&[55.02,55.58]&[83.11,84.05]&[84.10,84.76]    &[42.97,43.56]&[60.27,61.04]&[89.65,90.05]&[90.27,90.77]\\
		\multirow{2}{*}{Scaling}   &22.93&37.11&32.16&58.27	    &30.66&47.61&55.70&84.82	&34.83&53.53&84.34&90.89\\
		&[22.40,23.46]&[36.40,37.81]&[30.81,33.50]&[56.19,60.36]    &[30.44,30.89]&[47.12,48.11]&[53.70,57.71]&[84.53,85.12]    &[34.52,35.15]&[53.32,53.75]&[83.83,84.86]&[90.74,91.05]\\
		\multirow{2}{*}{Magnify}   &20.21&36.08&36.67&58.59	    &27.77&46.91&64.60&86.08	&31.65&54.90&87.02&91.11\\
		&[19.68,20.73]&[35.15,37.01]&[35.20,38.14]&[56.91,60.26]    &[27.48,28.05]&[46.08,47.74]&[61.87,67.33]&[85.74,86.42]   &[31.36,31.93]&[54.61,55.18]&[86.69,87.36]&[90.87,91.35]\\
		\multirow{2}{*}{Inverting} &26.51&38.89&38.74&59.89	    &35.44&49.24&81.81&85.37	&39.48&55.60&89.79&91.05\\
		&[26.00,27.00]&[38.04,39.74]&[36.90,40.58]&[58.37,61.41]    &[35.05,35.83]&[48.87,49.60]&[81.27,82.35]&[85.10,85.63]    &[39.15,39.82]&[55.38,55.83]&[89.55,90.03]&[90.88,91.21]\\
		\multirow{2}{*}{Reversing} &45.94&59.92&57.98&65.83	    &58.11&72.14&82.94&83.84	&62.79&76.80&90.29&90.53\\
		&[44.85,47.02]&[58.78,61.05]&[56.87,59.09]&[64.27,67.39]    &[57.75,58.47]&[71.81,72.47]&[82.43,83.44]&[83.39,84.30]    &[62.54,63.04]&[76.66,76.95]&[90.11,90.46]&[90.27,90.79]\\
		\multirow{2}{*}{Resampling(ours)}  &\textbf{69.69}&\textbf{68.60}&\textbf{70.97}&\textbf{69.20}	    &\textbf{84.84}&\textbf{78.51}&\textbf{87.58}&\textbf{85.84}    &\textbf{88.66}&\textbf{82.07}&\textbf{92.29}&\textbf{91.32}\\
		&[68.90,70.49]&[67.66,69.55]&[70.14,71.80]&[67.87,70.52]    &[84.61,85.07]&[78.21,78.81]&[87.34,87.81]&[85.65,86.02]    &[88.49,88.83]&[81.82,82.32]&[92.10,92.47]&[91.12,91.51]\\
		\bottomrule
		
	\end{tabular}
}

\end{table*}
\begin{table*}[]

\scriptsize

\centering

\caption{The statistical analysis results of the resampling augmentation method with other methods about the Wilcoxon signed-rank test on contrastive learning tasks}

\label{Tab06}

 \setlength{\tabcolsep}{2mm}{
	\begin{tabular}{ccccccccccccc}
		
		\toprule
		
		\multirow{3}{*}{} 
		&\multicolumn{4}{c}{1\%} &\multicolumn{4}{c}{10\%} &\multicolumn{4}{c}{60\%}\\
		\cmidrule(r){2-5}\cmidrule(r){6-9}\cmidrule(r){10-13}
		& \multicolumn{2}{c}{Linear evaluation} & \multicolumn{2}{c}{Fine-tuned} 
		& \multicolumn{2}{c}{Linear evaluation} & \multicolumn{2}{c}{Fine-tuned}
		& \multicolumn{2}{c}{Linear evaluation} & \multicolumn{2}{c}{Fine-tuned}\\
		
		\cmidrule(r){2-3} \cmidrule(r){4-5}\cmidrule(r){6-7} \cmidrule(r){8-9}
		\cmidrule(r){10-11} \cmidrule(r){12-13}
		
		&SimCLR. & MoCo.  &SimCLR. &MoCo.&SimCLR. & MoCo.  &SimCLR. &MoCo.
		&SimCLR. & MoCo.  &SimCLR. &MoCo.\\
		\midrule
		UCI-HAR(Sup.) &s- &s+ &s+ &s+   &s-&s-&s+&s+    &s-&s-&s+&s+\\
		Noise         &s+ &s+ &s+ &s+   &s+&s+&s+&s+    &s+&s+&s-&s-\\
		Rotated       &s+ &s+ &s+ &s+   &s+&s+&s+&s+    &s+&s+&- &+\\
		Scaling       &s+ &s+ &s+ &s+   &s+&s+&s+&s+    &s+&s+&+ &s-\\
		Magnify       &s+ &s+ &s+ &s+   &s+&s+&s+&s+    &s+&s+&- &+\\
	    Inverting     &s+ &s+ &s+ &s+   &s+&s+&+ &+     &s+&s+&+ &s+\\
		Reversing     &s+ &s+ &s+ &s+   &s+&s+&s+&s+    &s+&s+&- &-\\
		\midrule
		MotionSense(Sup.)   &s-&s-&s+&s+    &s-&s-&s+&+     &s-&s-&s+&s+\\
		Noise               &s+&s+&s+&s+    &s+&s+&s+&+     &s+&s+&s+&+ \\
		Rotated             &s+&s+&s+&s+    &s+&s+&s+&+     &s+&s+&s+&s+\\
		Scaling             &s+&s+&s+&s+    &s+&s+&s+&-     &s+&s+&s+&+ \\
		Magnify             &s+&s+&s+&s+    &s+&s+&s+&-     &s+&s+&s+&+ \\
	    Inverting           &s+&s+&s+&s+    &s+&s+&s+&s+    &s+&s+&s+&s+\\
		Reversing           &s+&+ &+ &+     &s+&s+&+ &s+    &s+&s+&s+&s+\\
		\midrule
		USC-HAD(Sup.)    &s+&s+&s+&s+    &s-&s-&s+&+     &s-&s-&s+&s+\\
		Noise           &s+&s+&s+&s+    &s+&s+&s+&s+    &s+&s+&s+&s+\\
		Rotated         &s+&s+&s+&s+    &s+&s+&s+&s+    &s+&s+&s+&s+\\
		Scaling         &s+&s+&s+&s+    &s+&s+&s+&s+    &s+&s+&s+&s+\\
		Magnify         &s+&s+&s+&s+    &s+&s+&s+&s+    &s+&s+&s+&+ \\
	    Inverting       &s+&s+&s+&s+    &s+&s+&s+&s+    &s+&s+&s+&+ \\
		Reversing       &s+&s+&s+&s+    &s+&s+&s+&s+    &s+&s+&s+&s+\\
		\bottomrule
	\end{tabular}
}
\begin{tablenotes}
  	\centering
    \footnotesize
    \item All non-parametric statistical hypothesis test results have been shown here at the 0.05 level of significance. These signs "+", "-", "s+", "s-" indicate that the resampling method is insignificantly better than, insignificantly worse than, significantly better than, and significantly worse than other augmentation methods, respectively.
  \end{tablenotes}
\end{table*}

\par
In linear evaluation, the experiment results show that resampling augmentation methods outperformed all state-of-the-art methods. In particular, with 1\% labeled data, the resampling augmentation method was improves significantly on SimCLRHAR and MoCoHAR, outperforming the best method by 4.27\% and 8.26\% on the UCI-HAR, 9.46\% and 0.14\% on the MotionSensor, 23.75\% and 8.68\% on the UCI-HAR, respectively. However, most augmentation methods are significantly worse than supervised learning in the statistical analysis results.
Only in the MoCoHAR framework did the resampling augmentation methods outperform supervised learning at 1\% labeled data on UCI-HAR and USC-HAD datasets. The difference between the best and the worst results is approximately 33\% on USC-HAD dataset, which shows that contrastive learning is very much focused on the selection of augmentation methods.
With 60\% labeled data, the linear evaluation of the contrastive learning model is not as good as supervised learning. This result is caused by the fact that supervised learning with a clear classification task under a large proportion of labeled data is better in classification than contrastive learning, where the task is to learn a better sample representation.
\par
In the fine-tuning evaluation, the contrastive learning models that correspond to most of the augmentation methods outperform the supervised learning models for each proportion of the labeled data. The resampling augmentation methods are optimal on both SimCLRHAR and MoCoHAR for both 1\% and 10\% labeled data on UCI-HAR and USCHAD datasets. However, at 60\% labeled data, the resampling method was found to be insignificantly improved by calculating the 95\% confidence limits and non-parametric statistical analysis. The resampling method does not outperform other augmentation methods in the UCI-HAR dataset, but the difference is insignificant.
\par
Overall, the resampling augmentation method performs more significantly with a small amount of labeled data and can also achieve a insignificant improvement with a large amount of labeled data, regardless of whether the contrastive learning framework uses SimCLRHAR or MoCoHAR. This finding is in accordance with the feature that contrastive learning can perform significantly with a small amount of labeled data in fine-tuning.
In addition, MoCoHAR performs well relative to SimCLRHAR in linear evaluation protocols and larger batch size environments. Through the above discussion, it was demonstrated that the resampling augmentation method could indeed generate new sample data that represent the raw sample features as well as improve the performance of contrastive learning.
\subsection{Comparison with State-of-the-art augmentation methods based on the TPN backbone network}
The work \cite{tang2020exploring}  is the first time to apply contrastive learning to HAR, and it uses TPN \cite{saeed2019multi} as the backbone network to evaluate performance on MotionSensor data. It pretrained in a modified SimCLR framework, conducted a combined augmentation study to explore the best augmentation method, and it was experimentally concluded that the best augmentation method is to use rotated for both augmentations. We used the same experiment setup according to the work \cite{tang2020exploring}, with the same experiment parameters, where the activity data of five random subjects were chosen for the test set and the activity data of the remaining subjects were used for the training set. The evaluation process used a five-fold cross-validation to ensure that each subject's activity data is available as a test set. The experiment results are shown in Table \ref{Tab07}.
\par
The experiment results show that the resampling augmentation method outperforms the previous best method rotated in linear evaluation. But and the resampling method insignificantly outperforms the rotated methods by calculating the 95 confidence limits in fine-tuning evaluation. We analyze the work \cite{tang2020exploring} divides a large proportion of the training set, leading to an easier training process, making the classification performance difference insignificant. In conclusion, this finding demonstrates that the resampling augmentation method can also have better results under the TPN backbone network.

\begin{table}
  \centering
  \caption{TPN backbone network on MotionSensor dataset}
  \label{Tab07}
  \begin{tabular}[]{ccc}
    \toprule
      &Linear evaluation &Fine-tuned\\
    \midrule
   \multirow{2}{*}{Rotated \cite{tang2020exploring}}   &85.67  &91.41\\
   &[82.55,88.79]&[90.44,92.38]\\
    \multirow{2}{*}{Resampling(ours)}  &\textbf{89.50}  &\textbf{92.10}\\
     &[88.42,90.57]&[89.36,94.84]\\
    \bottomrule
  \end{tabular}
\end{table}
\subsection{Combinations of Augmentation Methods}
\begin{figure*}[htbp]
  \centering

\subfigure[Linear evaluation for SimCLRHAR]{
\begin{minipage}[t]{0.5\linewidth}
  \centering
  \includegraphics[width=3in]{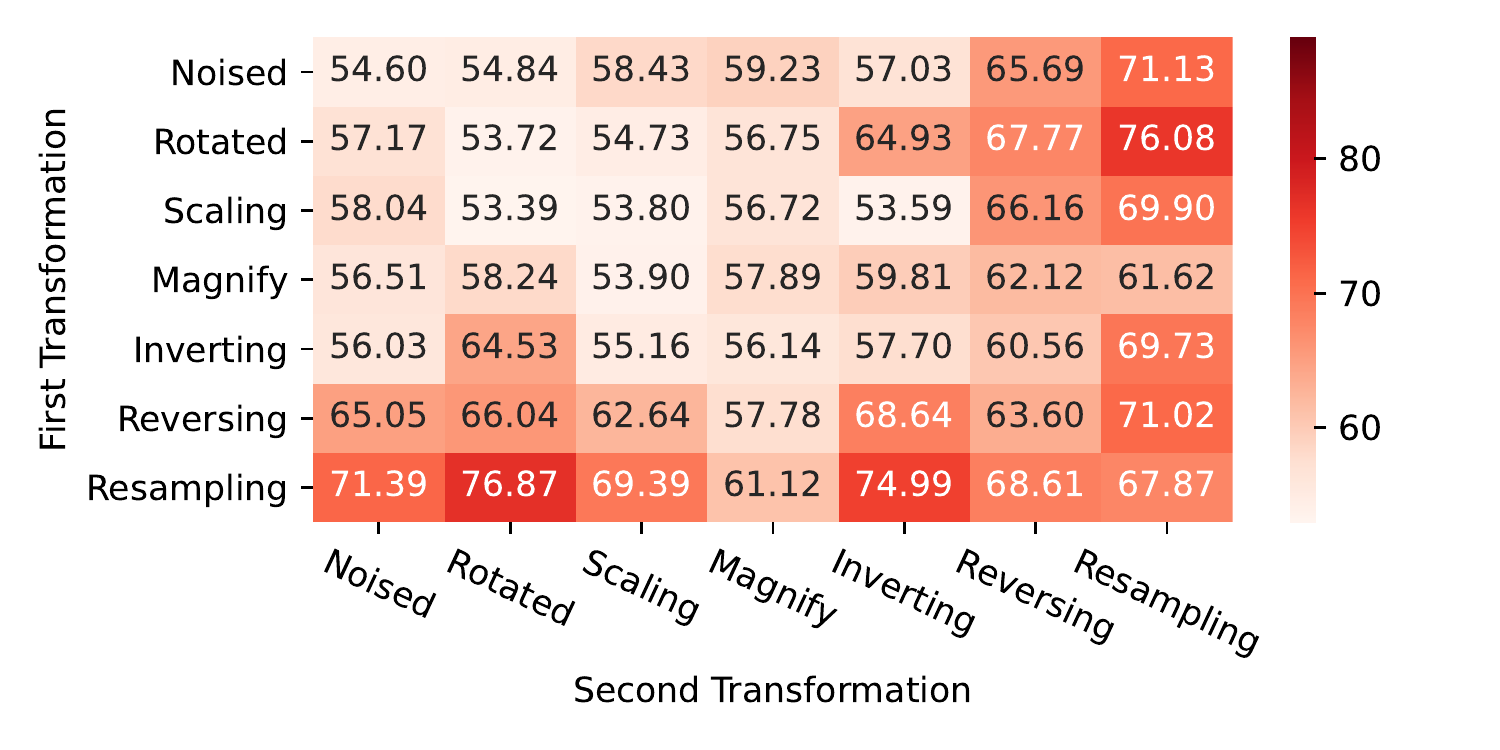}
\end{minipage}%
}%
\subfigure[Fine-tune for SimCLRHAR]{
\begin{minipage}[t]{0.5\linewidth}
  \centering
  \includegraphics[width=3in]{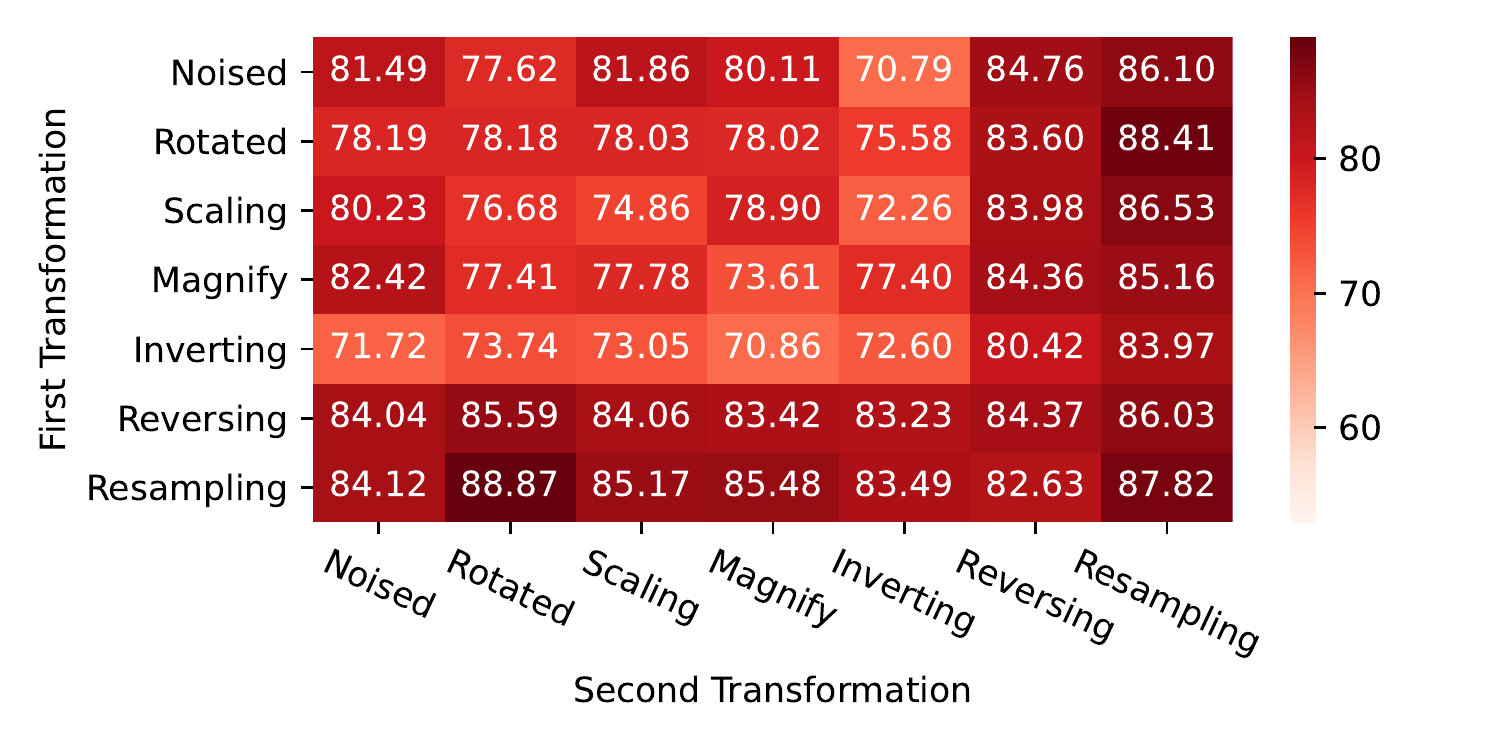}
\end{minipage}%
}%

\subfigure[Linear evaluation for MoCoHAR]{
\begin{minipage}[t]{0.5\linewidth}
  \centering
  \includegraphics[width=3in]{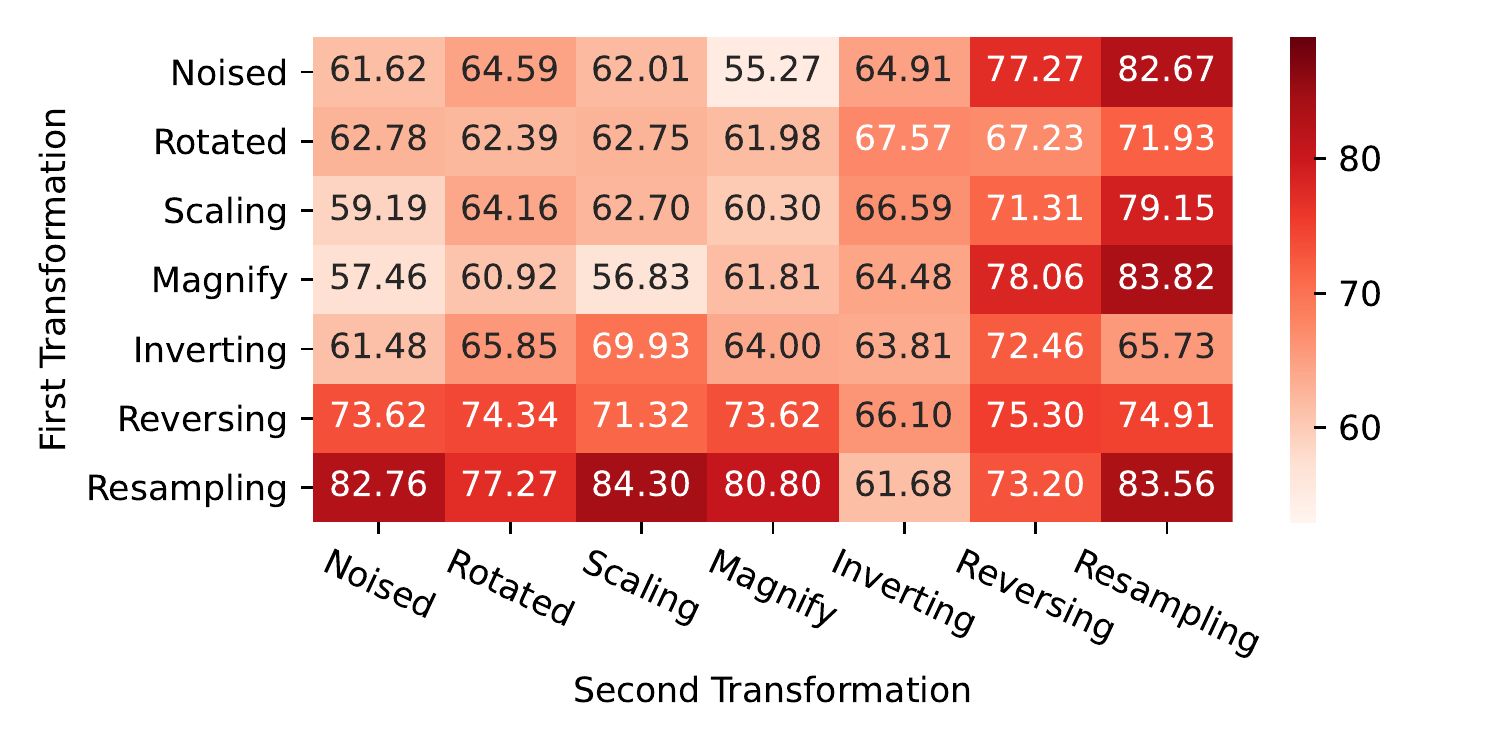}
\end{minipage}
}%
\subfigure[Fine-tune for MoCoHAR]{
\begin{minipage}[t]{0.5\linewidth}
  \centering
  \includegraphics[width=3in]{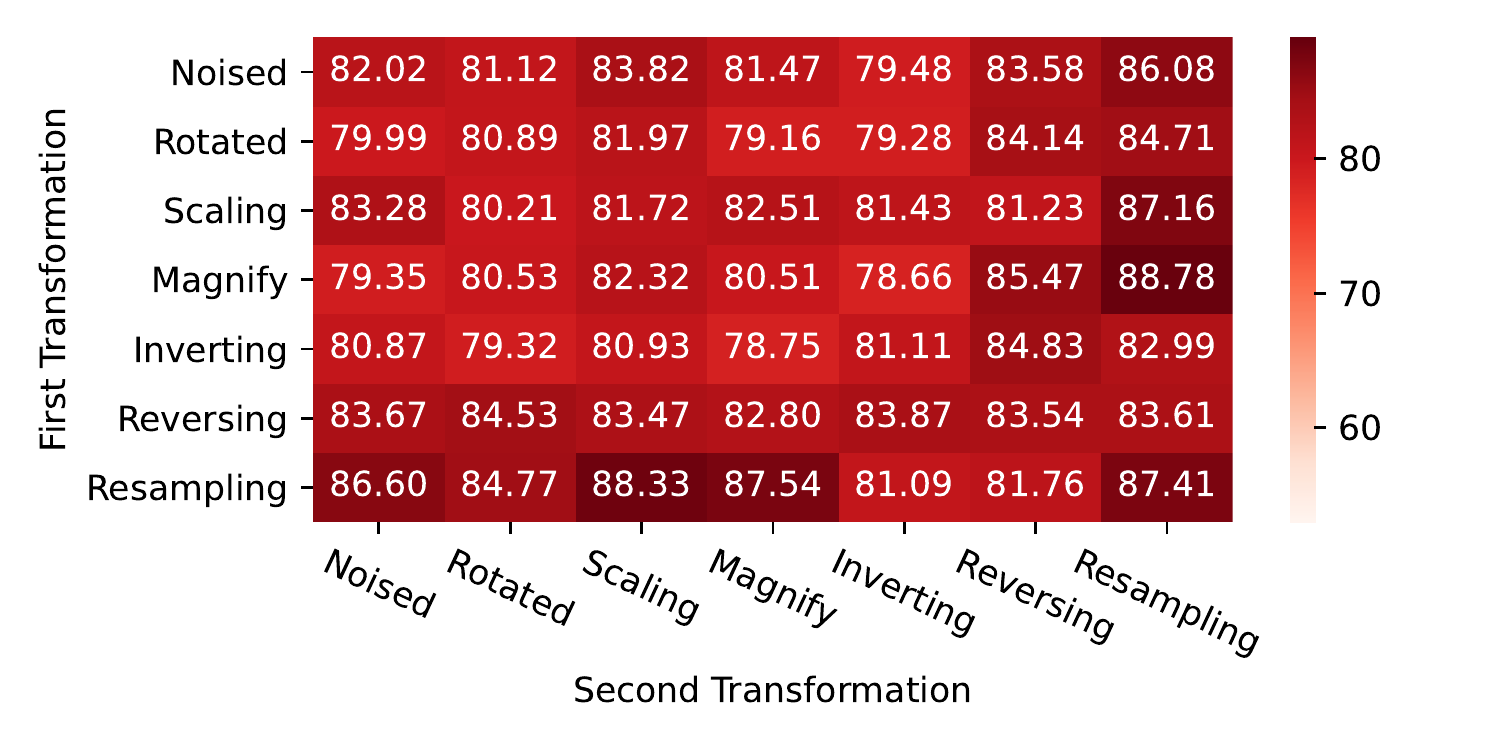}
\end{minipage}
}%
  \centering
\caption{Study on Combinatorial Augmentation Method. Diagonal Elements Use Individual Augmentation Methods.}
  \label{fig:4}
\end{figure*}

To explore the performance of different combinations of augmentation methods on contrastive learning, we use SimCLRHAR and MoCoHAR as a contrastive learning framework and use a combination of augmentation in the data augmentation phase. To compare more clearly with different combinations of augmentation methods, we do not take augmentation in the augmentation phase for the first branch, we use the raw samples, and we use a specific combination of augmentation methods for the second branch. This experiment uses 1\% labels as the training set for the downstream classification task, which is validated on the UCI-HAR dataset. The experiment results are shown in Fig. \ref{fig:4}.

\par The experiment results show that the combination of augmentation methods under the SimCLRHAR framework did not outperform supervised learning in the linear evaluation, but the combination of resampling and rotation was close to supervised learning and outperformed the best individual augmentation of resampling by 8.97\%. On the MoCoHAR framework, there are some combinations of augmentations that outperform supervised learning, the highest being the combination of resampling and scaling, which outperforms supervised learning by 6.36\% and outperforms the best individual augmentation by 0.74\%. In the fine-tuning evaluation, there are some combinations of augmentation methods under SimCLRHAR that outperform supervised learning, the best being the combination of resampling and rotation, outperforming supervised learning by 10.93\% and outperforming individual augmentation by 1.05\%. There are also some combinations of augmentation methods under the MoCoHAR framework that outperform supervised learning, the best being the combination of magnification and resampling, which outperforms supervised learning by 10.84\% and outperforms individual augmentation  by 1.37\%. All of the above combinations of optimal augmentation methods have resampling, which shows that resampling augmentation methods play an important role in combinatorial augmentation. It is also demonstrated that the combined augmentation is superior to using only individual augmentation in the SimCLRHAR and MoCoHAR frameworks.

\subsection{A Study of Batch Size}
\begin{figure*}[htbp]
	\centering
\subfigure[Linear evaluation in the UCI-HAR dataset, using 1\% labeled data]{
\begin{minipage}[t]{0.5\linewidth}
			\centering
			\includegraphics[width=3in]{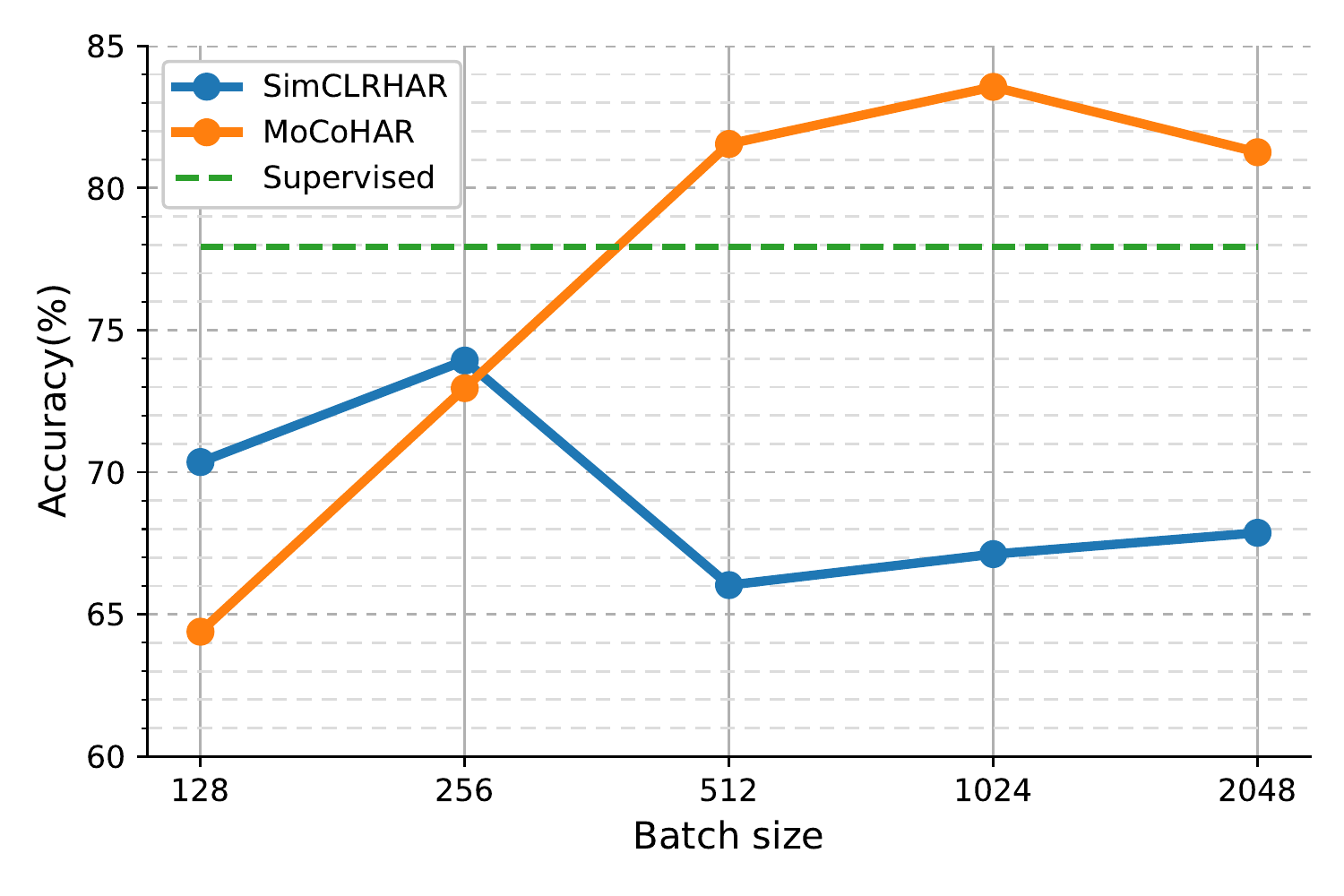}
\end{minipage}%
}%
\subfigure[Linear evaluation in the MotionSense dataset, using 1\% labeled data]{
\begin{minipage}[t]{0.5\linewidth}
			\centering
			\includegraphics[width=3in]{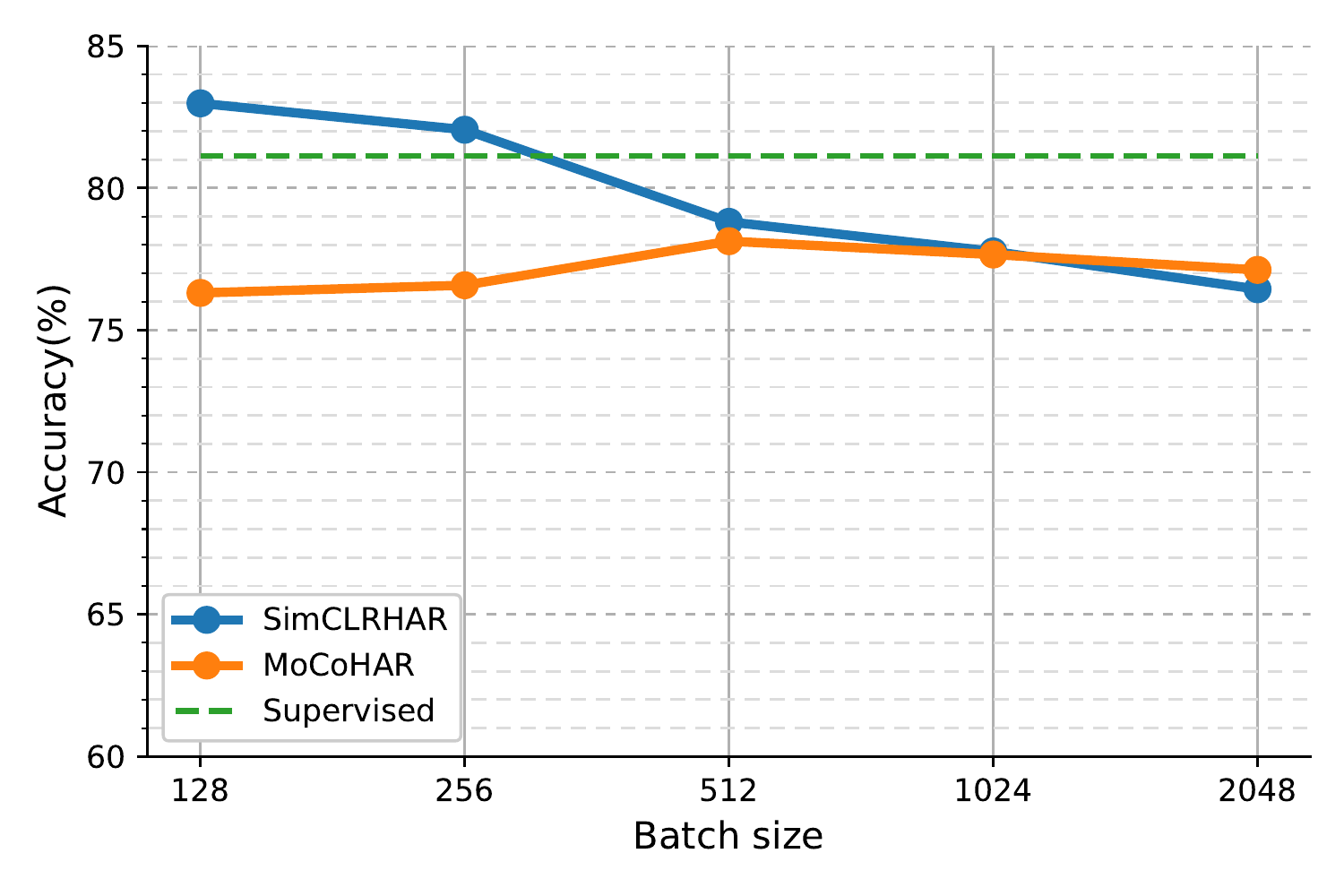}
\end{minipage}%
}
\caption{Impact of Batch Size on Contrastive Learning Framework (Mean F1-score). The supervised curve does not participate in the change of batch size.}
	\label{fig:5}
\end{figure*}

To explore the sensitivity of the contrastive learning framework to batch size, we trained two contrastive learning frameworks, SimCLRHAR and MoCoHAR, with different batch sizes in the pre-training phase and used a linear evaluation protocol to validate the classification performance after randomly selecting 1\% of the labeled data as the training set and the remainder as the test set. The experiment results are shown in Fig. \ref{fig:5}.
\par
SimCLRHAR is more effective at smaller batch sizes, while MoCoHAR is more effective at larger batch sizes. In the field of computer vision, MoCo is more effective than SimCLR at smaller batch sizes. We analyze the contrary  result produced here suggests that the sensitivity of sensor data to batch size under contrastive learning is not the same as that of image data.

\section{Conclusions}
In this paper, we propose a resampling sensor data augmentation method that simulates changing the sampling frequency. In addition, we extended MoCo for HAR to include a new resampling data augmentation  and  DeepConvLSTM  encoder,  which  is  called MoCoHAR. SimCLRHAR and MoCoHAR as two contrastive learning environments to evaluate the resampling data augmentation.
We conducted experiments on UCI-HAR, MotionSensor and USC-HAD, using multiple proportions of labeled data as the training set. The experiment results show that the resampling augmentation method outperforms all SOTA augmentation methods in both supervised learning and contrastive learning under a small amount of labeled data. In linear evaluation, with 1\% labeled data, the resampling augmentation method was improves significantly on SimCLRHAR and MoCoHAR, outperforming the best method by 4.27\% and 8.26\% on the UCI-HAR, 9.46\% and 0.14\% on the MotionSensor, 23.75\% and 8.68\% on the UCI-HAR, respectively. MoCoHAR performs well relative to SimCLRHAR in linear evaluation protocols and larger batch size environments. Finally, we also studied the performance of batch size and combined augmentation on contrastive learning. 
\par
In addition, there are some shortcomings in this paper. The performance improvement of this resampling augmentation method is insignificant with a large amount of labeled data. Also, we found that the inference speed of the resampling augmentation method is longer compared to other methods during the experimental phase. We will further improve the performance and inference speed of the resampling augmentation method in the future. We will also start with real-life applications of contrastive learning. For example, we use contrastive learning to analyze and train models on unlabeled data collected in any environment, so that the models in the new environment can perform well in activity recognition with only a small amount of labeled data  fine-tuning.


\par


%



\section*{Acknowledgements}
This work is supported by the National Natural Science Foundation of China (61872038, 62006110).


\ifCLASSOPTIONcaptionsoff
\newpage
\fi



\bibliographystyle{IEEEtran}
\bibliography{main}

%

%
%

%

\begin{IEEEbiography}[{\includegraphics[width=1in,height=1.25in,clip,keepaspectratio]{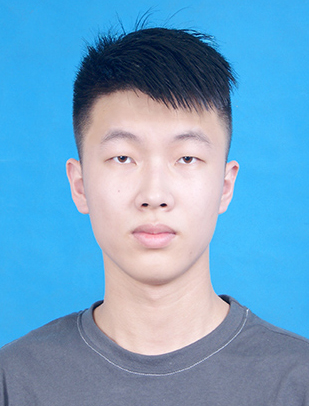}}]{Jinqiang Wang}
received his B.E. degree from Henan Normal University in 2020. He is currently a M.S. student in the School of Computer Science, University of South China. His research interests include intelligent perception and pattern recognition.
\end{IEEEbiography}


\begin{IEEEbiography}[{\includegraphics[width=1in,height=1.25in,clip,keepaspectratio]{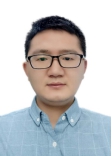}}]{Tao Zhu}
received the Ph.D. degree from University
of Science and Technology of China in 2015 and the BE
degree from Central South University in 2009. Then,
he worked as a post-Ph.D. and a lecturer in School of
Computer and Communication Engineering, University
of Science and Technology Beijing. Currently, he is
with University of South China. His research interests
include Evolutionary Computation and Internet of Things.
\end{IEEEbiography}

\begin{IEEEbiography}[{\includegraphics[width=1in,height=1.25in,clip,keepaspectratio]{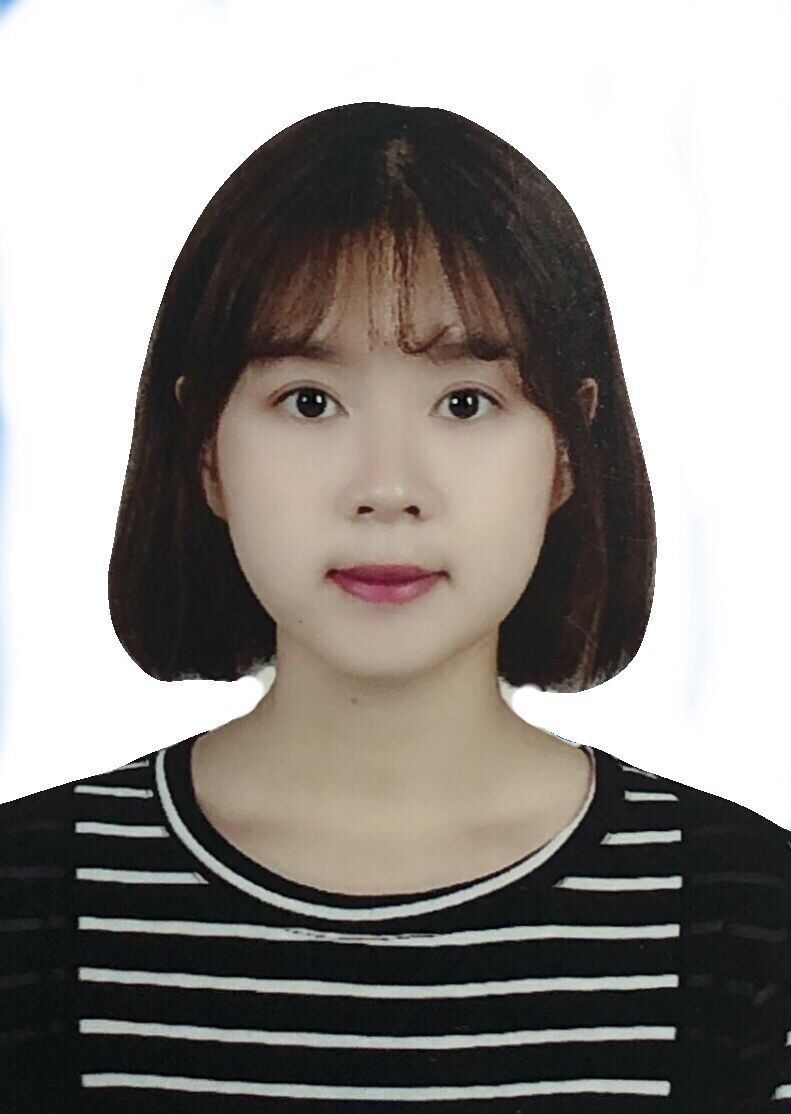}}]{Jingyuan Gan}
is currently a B.E. student in the School of Computer Science, University of South China. Her research interests include intelligent perception and pattern recognition.
\end{IEEEbiography}
\begin{IEEEbiography}[{\includegraphics[width=1in,height=1.25in,clip,keepaspectratio]{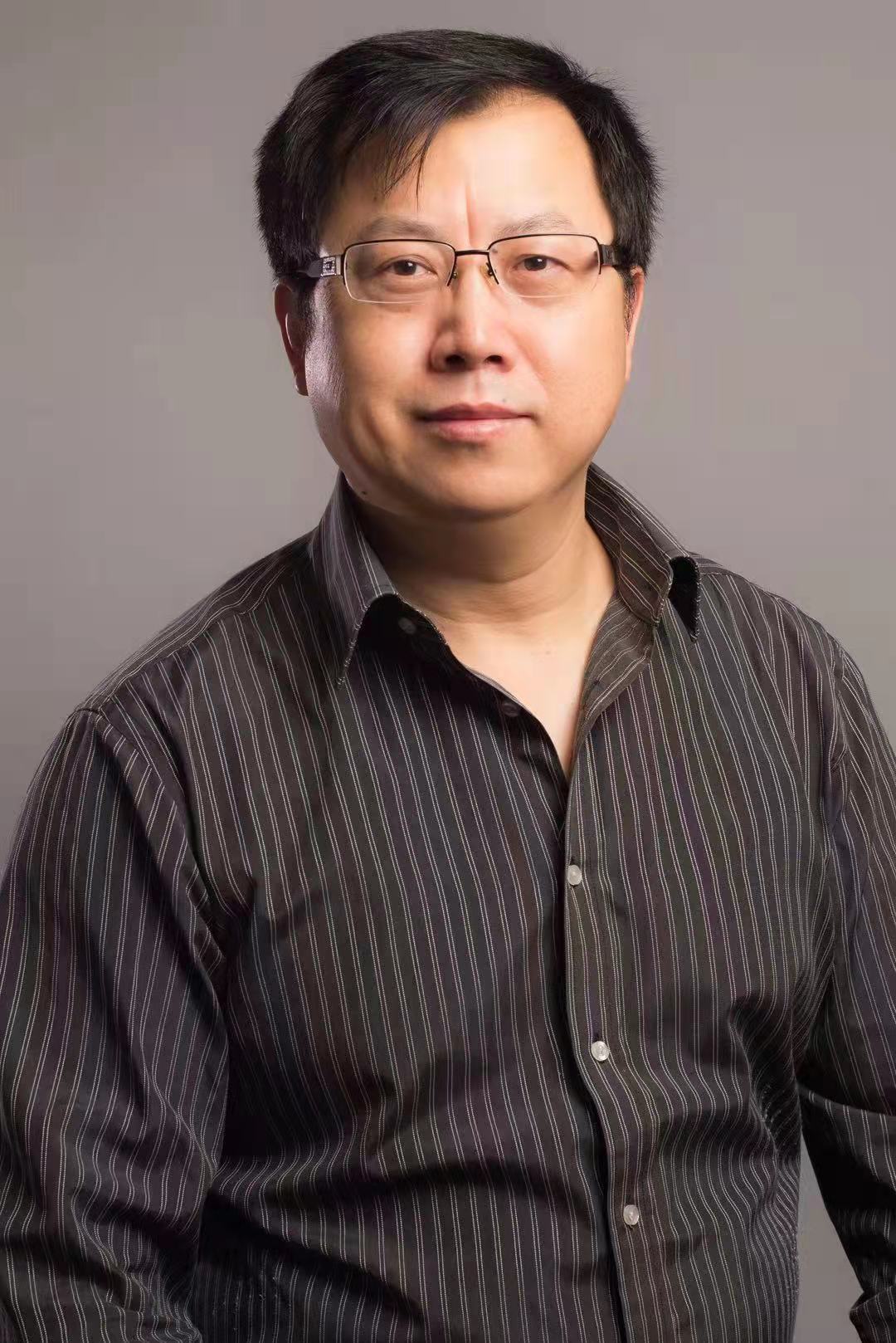}}]{Liming Luke Chen}
is Professor of Data Analytics in the School of Computing, Ulster University, UK. He received his BEng and MEng degrees at Beijing Institute of Technology, China, and DPhil on Computer Science at De Montfort University, UK. His current research interests include pervasive computing, data analytics, artificial intelligence, user-centred intelligent systems and their applications in smart healthcare and cyber security. He has published over 250 papers in the aforementioned areas. Liming is an IET Fellow and a Senior Member of IEEE.
\end{IEEEbiography}
\begin{IEEEbiography}[{\includegraphics[width=1in,height=1.25in,clip,keepaspectratio]{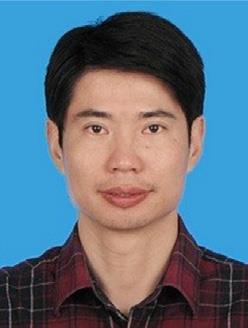}}]{Huansheng Ning}
received his B.S. degree from Anhui
University in 1996 and his Ph.D. degree from Beihang
University in 2001. He is currently a Professor and Vice
Dean with the School of Computer and Communication
Engineering, University of Science and Technology
Beijing and China and Beijing Engineering Research
Center for Cyberspace Data Analysis and Applications,
China, and the founder and principal at Cybermatics
and Cyberspace International Science and Technology
Cooperation Base. He has authored several books and
over 70 papers in journals and at international conferences/
workshops. He has been the Associate Editor of IEEE Systems Journal
and IEEE Internet of Things Journal, Chairman (2012) and Executive Chairman
(2013) of the program committee at the IEEE international Internet of Things
Conference, and the Co-Executive Chairman of the 2013 International Cyber
Technology Conference and the 2015 Smart World Congress. His awards include
the IEEE Computer Society Meritorious Service Award and the IEEE Computer
Society Golden Core Member Award. His current research interests include
Internet of Things, Cyber Physical Social Systems, electromagnetic sensing and
computing.
\end{IEEEbiography}

\begin{IEEEbiography}[{\includegraphics[width=1in,height=1.25in,clip,keepaspectratio]{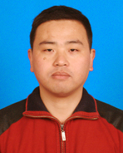}}]{Yaping Wan}
required Ph.D. degree from Huazhong University of Science and Technology. His research interests include big data causal inference.
\end{IEEEbiography}




\end{document}